\documentclass[fleqn,usenatbib]{mnras}

\usepackage{newtxtext,newtxmath}

\usepackage[T1]{fontenc}

\usepackage{graphicx}	
\usepackage{amsmath}	
\usepackage{gensymb}
\usepackage{comment}

\usepackage{booktabs}
\usepackage{longtable, lipsum}
\usepackage{bm}
\usepackage{hyperref}

\title[Galactic magnetism with pulsars]{The Thousand-Pulsar-Array programme on MeerKAT -- XVI. Mapping the Galactic magnetic field with pulsar observations}

\author[L. S. Oswald et al.]{L. S. Oswald$^{1,2,3}$\thanks{E-mail: lucy.oswald@physics.ox.ac.uk (LSO)}, P.~Weltevrede$^{4}$, B.~Posselt$^{1,5}$, S.~Johnston$^{6}$, A.~Karastergiou$^{1}$, M.~E.~Lower$^{6,7}$
\\
$^{1}$Department of Astrophysics, University of Oxford, Denys Wilkinson Building, Keble Road, Oxford OX1 3RH, UK\\
$^{2}$Magdalen College, University of Oxford, Oxford OX1 4AU, UK\\
$^{3}$School of Physics \& Astronomy, University of Southampton, Southampton SO17 1BJ, UK\\
$^{4}$Jodrell Bank Centre for Astrophysics, Department of Physics and Astronomy, University of Manchester, Manchester M13 9PL, UK\\
$^{5}$Department of Astronomy \& Astrophysics, Pennsylvania State University, 525 Davey Lab, University Park, PA 16802, USA\\
$^{6}$Australia Telescope National Facility, CSIRO, Space and Astronomy, PO Box 76, Epping, NSW 1710, Australia\\
$^{7}$Centre for Astrophysics and Supercomputing, Swinburne University of Technology, PO Box 218, Hawthorn VIC 3122, Australia\\
}

\date{Accepted 2025 March 31. Received 2025 February 28; in original form 2024 July 15}

\pubyear{2025}

\begin{document}
\label{firstpage}
\pagerange{\pageref{firstpage}--\pageref{lastpage}}
\maketitle

\begin{abstract}

Measuring the magnetic field of the Milky Way reveals the structure and evolution of the galaxy. Pulsar rotation measures (RMs) provide a means to probe this Galactic magnetic field (GMF) in three dimensions. We use the largest single-origin data set of pulsar measurements, from the MeerKAT Thousand-Pulsar-Array, to map out GMF components parallel to pulsar lines of sight. We also present these measurements for easy integration into the consolidated RM catalogue, RMTable. Focusing on the Galactic disk, we investigate competing theories of how the GMF relates to the spiral arms, comparing our observational map with five analytic models of magnetic field structure. We also analyse RMs to extragalactic radio sources, to help build up a three-dimensional picture of the magnetic structure of the galaxy. In particular, our large number of measurements allows us to investigate differing magnetic field behaviour in the upper and lower halves of the Galactic plane. We find that the GMF is best explained as following the spiral arms in a roughly bisymmetric structure, with antisymmetric parity with respect to the Galactic plane. This picture is complicated by variations in parity on different spiral arms, and the parity change location appears to be shifted by a distance of 0.15~kpc perpendicular to the Galactic plane. This indicates a complex relationship between the large-scale distributions of matter and magnetic fields in our galaxy. Future pulsar discoveries will help reveal the origins of this relationship with greater precision, as well as probing the locations of local magnetic field inhomogenities.

\end{abstract}




\section{Introduction}

The study of galaxy formation and evolution requires a detailed understanding of Galactic magnetic fields. From our vantage point on Earth, the Milky Way is the source of the most detailed information of Galactic structure, because we are able to view its Galactic magnetic field (GMF) in three dimensions. A key observational tracer of Galactic magnetism is pulsar radio emission: with just two straightforward measurements, the dispersion and rotation measures, we are able to estimate of the average magnetic field strength along the line of sight to the pulsar \citep{Manchester1972}.

Observational tracers of the GMF are varied and are sensitive to different components of the field, either parallel or perpendicular to the line of sight. Tracers of the perpendicular field component include synchrotron emission and all-sky dust polarization maps \citep[full summary by][]{Haverkorn2019} and starlight polarization \citep{Pelgrims2024}. Newly proposed techniques include using Fast Radio Bursts to map the parallel magnetic field component along the entire line of sight through the galaxy \citep{Pandhi2022}, and applying the Velocity Gradient Technique to HI data to obtain full 3D mapping \citep{Hu2023}. Two key techniques for obtaining the parallel component of the GMF along the line of sight involve measuring the strength of Faraday rotation from radio sources \citep[e.g.][]{OSullivan2023a}. Measurements of Faraday rotation to extragalactic radio sources (EGRS) have recently provided a detailed map of the full Faraday depth of the galaxy \citep{Hutschenreuter2022}. Combined with models and evidence of the Galactic electron column density, \cite{Hutschenreuter2023} then inferred an all-sky two-dimensional map of the magnetic field strength parallel component averaged on sightlines through the whole galaxy. Radio pulsars are also highly polarized sources to which Faraday depths can be measured, but unlike EGRS, they are unique in having two important attributes. First, dispersion of the pulsar signal provides an independent estimate of the electron column density, so that the average magnetic field strength itself can be estimated directly. Second, they are situated inside the galaxy, enabling the study of its three-dimensional internal structure, rather than the two-dimensional view of the sky.  

The physics behind this is as follows. As the polarized radio beam of the pulsar travels through the interstellar medium (ISM) from source to observer, it undergoes dispersion and Faraday rotation. The cumulative effects of these processes are described by integrating the electron column density and the magnetic field strength along the line of sight to calculate the dispersion measure (DM) and rotation measure (RM):
\begin{equation}
    \textrm{DM} = \int_{0}^{d} n_{e} dl
    \label{eq:DM}
\end{equation}
\begin{equation}
    \textrm{RM} = \frac{e^{3}}{8\uppi^{2}\upvarepsilon_{0}m_{e}^{2}c^{3}} \int_{0}^{d} n_{e} \bm{B}\cdot\bm{dl}
    \label{eq:RM}
\end{equation}
where $n_{e}$ is the Galactic electron number density distribution, $\bm{B}$ is the three-dimensional magnetic field vector, $\bm{dl}$ describes the line of sight vector from pulsar to observer, $d$ is the distance to the pulsar and the constant of proportionality in front of the RM integral is given in a form such that if the constants $e$, $\epsilon_{0}$, $m_{e}$ and $c$ are all measured in SI units, the combined constant of proportionality has units of inverse Tesla ($T^{-1}$). We can separate out the magnetic field component if we assume that $n_{e}$ and $\bm{B}$ are not correlated with each other and can be treated independently. This has the potential to either over or under-estimate the GMF by up to a factor of 3 if the assumption is incorrect \citep{Beck2003}, however, \cite{Seta2021} demonstrated that the assumption is valid on kpc scales. We may therefore obtain the component of $\bm{B}$ that lies parallel to the line of sight, $B_{\parallel}$, averaged along the path to the pulsar, as being proportional to the ratio of RM and DM: 
\begin{equation}
    \langle B_{\parallel}\rangle = \frac{\int_{0}^{d} n_{e} B_{\parallel}dl}{\int_{0}^{d} n_{e} dl} \propto \frac{\textrm{RM}}{\textrm{DM}}.
    \label{eq:Bpropto}
\end{equation}
Usual measurement units for the observables are cm$^{-3}$pc for the DM and rad~m$^{-2}$ for the RM. To obtain $\langle B_{\parallel}\rangle$ in units of $\upmu$G, we must therefore scale the integration constant in equation \ref{eq:RM} by $10^{-4}\left(\frac{1\textrm{pc}}{1\textrm{m}}\right)$, giving the following constant of proportionality for equation \ref{eq:Bpropto}: 
\begin{equation}
    \langle B_{\parallel}\rangle = \left( 10^{-4}\left(\frac{1\textrm{pc}}{1\textrm{m}}\right) \frac{e^{3}}{8\uppi^{2}\upvarepsilon_{0}m_{e}^{2}c^{3}} \right)^{-1} \frac{\textrm{RM}}{\textrm{DM}} \approx 1.232 \frac{\textrm{RM}}{\textrm{DM}}.
    \label{eq:aveB}
\end{equation}
Authors have used these measurements to test analytical models of the GMF in the plane \citep[e.g.][]{Noutsos2008b, Men2008}; to measure the scale height of the halo GMF \citep{Sobey2019}; and to probe GMF behaviour on both local and global length scales, including the Galactic spiral arm and inter-arm regions \citep{Han2018}.

Early models of the GMF structure include the concentric ring model \citep{Rand1989}, which is now largely superseded by the axisymmetric and bisymmetric spiral models where the magnetic field follows the Galactic spiral arms \citep[see][and references therein]{Beck1996}. In the spiral models, the direction of the field points along these spiral arms, either towards or away from the centre. For the axisymmetric model, the field direction follows the same pattern for all the spiral arms, whereas for the bisymmetric model the field direction can be followed into the centre along one spiral arm and out from the centre along another, so that overall the field direction alternates from one spiral arm to another \citep{Han2003}. In general, attempts to fit such models to pulsar observations have shown that they are unable to explain the complexity of the data \citep[e.g.][]{Noutsos2008b, Men2008, Nota2010}. This is due at least in part to local regions having a strong impact on the observed $\langle B_{\parallel}\rangle$ over and above the global GMF, but it may also be the case that simple models simply do not describe the full complexity of the field.

Newer GMF models have been developed and tested using alternative observations, predominantly synchrotron intensity and RMs to extragalactic radio sources. \cite{Jaffe2010} modelled the GMF on the plane, simulating synchrotron and EGRS observables and comparing these to existing observations. They sought to separate and constrain three components of the GMF: the coherent, random and ordered components, assuming a 2D logarithmic spiral model. \cite{VanEck2011} combined EGRS and pulsar measurements to model the disk field, concluding that the inner field followed the spiral arms and the outer field was azimuthal, with one region spiralling out from the centre in which the field is reversed. \cite{Jansson2012}, again focusing on synchrotron and EGRS, favoured a spiral disk field and, in particular, demonstrated evidence for a significant poloidal halo field component. 
\cite{Terral2017}, using Faraday depths to EGRS, modelled spiralling magnetic fields in the Galactic halo and found a slight preference for a bisymmetric structure. Meanwhile, \cite{Vallee2022} favoured an axisymmetric spiral model for which the pitch angle is somewhat decoupled from that of the spiral arms, and there exist reversals in field direction that form an annular shape. 
Most recently, \cite{Unger2023} combined EGRS and synchrotron measurements to test a range of models of the GMF, encompassing the disk field and both poloidal and toroidal halo fields. Their work supports the poloidal halo field contribution proposed by \cite{Jansson2012}. It also demonstrates that evidence for a large-scale spiral field in the disk remains inconclusive, with a local spiral spur fitting the data set comparably well. This is predominantly due to the lack of constraining ability of EGRS measurements in the plane, since they can only provide an average of the entire line of sight through the galaxy. The authors comment that pulsars could be used to help break this degeneracy, but that they in turn are constrained by the limited knowledge of their distances from Earth.

Galactic diffuse synchrotron emission is another important probe of the GMF. Recent works comparing diffuse emission to EGRS have included \cite{Ordog2019}, who found that the RMs from polarized extended emission track those from EGRS except where lines of sight cross regions with large amounts of HII. \cite{Erceg2022} used LOFAR images to reveal diffuse polarized emission morphology at low frequencies, and again found a correlation with the Faraday sky from EGRS. Field reversals have been studied in detail using diffuse emission, with \cite{Ordog2017} identifying a field reversal region which is diagonal with respect to the galactic plane, and \cite{Dickey2022} also identifying a field reversal which appears to shift above the Galactic plane with increasing Galactic radius. These studies reveal the complex relationships between GMF behaviour in the Galactic plane and in the halo, and \cite{Ordog2017} in particular emphasise the need for three-dimensional studies of the GMF to break degeneracies.

The most recent works to use pulsars to model the GMF are those of \cite{Han2018}, who compared pulsars with background sources to investigate large scale field reversals and developed a model for the disk magnetic field, and \cite{Xu2022}, who focused on directions of the field in the spiral arms and inter-arm regions in the first Galactic quadrant only. Since then, new works have constrained rotation measures to new pulsars \citep[e.g.][]{Ng2020} and to investigate the GMF of the Galactic halo through pulsar measurements in the globular cluster 47 Tuc \citep{Abbate2020a}. But no new attempts have been made to fit global GMF models to pulsar measurements of $\langle B_{\parallel}\rangle$.

The main limitation of using pulsars to constrain the GMF so far has been the coarse sampling of lines of sight through the galaxy that they provide, due to limited numbers of sources. As more pulsars have been discovered, the coverage of the map of $\langle B_{\parallel}\rangle$ in the galaxy has improved. Publications mapping the GMF with pulsars have progressed from using 19 RMs \citep{Manchester1972} to 1,222 RMs \citep{Ng2020} (1,167 catalogue values and 55 new measurements at time of publication). Now, the Thousand-Pulsar-Array (TPA) project presents RM measurements resulting from 19,697 observations of 1,097 pulsars, the largest single cohesive data-set of this sort to date, which includes 254 pulsars for which new RM measurements were presented for the first time by \cite{Posselt2023}. Using these, combined with 741 previously recorded measurements from the pulsar catalogue \citep[][version 2.1.1]{Manchester2005} we create a catalogue of measurements to 1,838 pulsars in total with which to probe the three-dimensional internal structure of the GMF.

In Section \ref{sec:methods} we give details of the observations used and the methods for calculating dispersion and rotation measures, along with information about extragalactic radio source data used for comparison. We also summarise the modelling approaches we take for estimating pulsar distances, and for comparing our observations to analytic models of the Galactic magnetic field. We present the results of our measurements in Section \ref{sec:results} in tables (provided in full machine-readable format in the Supplementary Material) and a sky map of the average magnetic field along the line of sight to each pulsar. We also present the full collection of RMs formatted according to the RMTable format, to enable straightforward integration with the universal catalogue of rotation measures from radio sources, RMTable2023 \citep{VanEck2023}. In Section \ref{sec:galplaneresults} we apply these results to a study of the magnetic field in the Galactic plane, investigating spiral configurations and comparing results with previous studies and with measurements of Faraday rotation to extragalactic radio sources. We follow this up by performing a quantitative comparison of our data set with a set of representative analytic magnetic field models in section \ref{sec:analyticmodels}. Further perspectives on the outcome of this modelling, the current limitations and prospects for the future are discussed in Section \ref{sec:discussion}, and Section \ref{sec:conclusions} summarises the key conclusions.

\section{Observations, measurements and modelling methods}
\label{sec:methods}

\subsection{Pulsar observations and rotation measures}
\label{sec:RMmethods}

We consolidate here a catalogue of the largest cohesive set of pulsar RM and DM measurements made with a single instrument, which have specifically been tailored to be the most representative sample of measurements of the large-scale GMF. 
Our analysis here is focused on observations of radio pulsars from the Thousand-Pulsar-Array project, part of the MeerTime Large Survey Project on MeerKAT \citep{Bailes2020b, Johnston2020}. We conducted a census of 1,170 pulsars and then continued follow-up monitoring observations on a monthly basis for 597 pulsars \citep{Keith2024}. Details of the observing strategy are given in \cite{Song2020} and the full data reduction strategy is presented in \cite{Posselt2023}. Key parameters of the observations are as follows. The pulsars were observed using the MeerKAT L-band receiver, from 896 to 1671 MHz with 928 frequency channels. The data are folded into 8-s sub-integrations and we record 1024 phase bins per pulse period. We combine this data set with existing measurements to other pulsars obtained from the pulsar catalogue {\sc{psrcat}} \citep[][version 2.1.1]{Manchester2005}. For all consultations of {\sc psrcat}, we make use of {\sc psrqpy} \citep{Pitkin2018}.

For this work we made use of the long time baseline of our observations to account for time-varying fluctuations in RM measurements, which we observe to be predominantly caused by the varying effect of the Earth's ionosphere, and by the impact of radio frequency interference (RFI) on individual measurements, although the impact of changing lines of sight from the Earth to the pulsar, particularly through its local environment, also has an important impact. It is by accounting for these time-varying fluctuations in RM that we are able to present the most representative sample of RMs suitable for GMF modelling. This is because, by measuring average RM values over the course of the observing time period, we ensure that these values are representative of the Galactic contribution to Faraday rotation, rather than being unduly affected by temporary fluctuations from sources more local either the pulsar or the Earth, in the way that a census observation RM may be. Our method for doing this is as follows.

We performed a measurement of the RM for every pulsar observation in the data-set of 24,108 observations of 1,242 pulsars, taken between 12th February 2019 and 15th May 2023, using the RM synthesis technique as described by \cite{Brentjens2005} and implemented in {\sc psrsalsa} \citep{Weltevrede2016}. Further details of this are given in \cite{Keith2024}. The RM measurement is based on a fit of a sinc-like function to the output RM synthesis power of trial RMs covering 20 rad~m$^{-2}$ around the expected RM. To avoid bias, after finding the most likely fit, we then repeat the measurement covering a 400 rad~m$^{-2}$ range, after centring the search range at the fitted RM. The implementation of RM synthesis on every pulsar observation is a largely automated process, but there exist observations within the sample for which the pulsar is not visible, either due to intermittency of the pulsar or due to excessive RFI. We therefore performed a by-eye rejection of pulse profiles for which the signal to noise ratio pulse profile itself was too low, since for these observations an RM synthesis fit will be meaningless. This process removed 1,799 observations of 26 pulsars. Further details of this approach are given by \cite{Keith2024}, who presented the full RM time series measurements for 597 pulsars from this sample. Our work here makes use of those same time series, but extends the RM measurement process to incorporate all TPA pulsar observations, including those without long-term follow-up monitoring.

Next, we performed a more in depth analysis of individual observations for which the RM could not be accurately constrained: these cases were predominantly for observations with low signal-to-noise ratio or those adversely affected by RFI or strong scintillation. In order to do this efficiently for the remaining 22,309 measurements of 1,216 pulsars that required analysis, we used the following metric. We removed all measurements for which the error on the measured RM was greater than 7~$\textrm{rad}~$m$^{-2}$. We chose this cut-off as being roughly 5 exponential decay-lengths on our distribution of RM errors, so that around 10\% of the observations are removed (2,587 observations removed; 19,722 observations of 1,097 pulsars remaining). For the remaining observations, we subtracted the estimated ionospheric contribution to the RM, identified using the software {\sc ionfr} \citep{Sotomayor-Beltran2013}. We then identified and removed RM values which were clear outliers relative to the distributions of RM measurements per pulsar: we defined this as measurements lying more than 10 units away from the median RM value for pulsars with more than 10 observations (a further 25 observations removed). This left 19,697 observations of 1,097 pulsars: 329 pulsars having only one RM measurement and the remaining 768 having an average (median) of 49 measurements contributing. We then calculated the median and median absolute deviation (MAD) RM per pulsar. These average measurements encompass any remaining uncertainty due to insufficiently modelled ionospheric fluctuations or small scale measurement deviations due to changing lines of sight or instrument noise, and therefore represent the largest uniform and representative survey of the average contribution of the GMF to pulsar rotation measures to date. 

Checking the extent to which our removal of bad observations affects our median calculation, we find that it has a non-zero effect for 220 pulsars, usually altering RMs by up to 5 units, with the biggest alteration being from 86.2 to 64.2~$\textrm{rad}~$m$^{-2}$ for PSR~J1835$-$0643 (two out of its three observations were dropped due to excessively large measurement uncertainties). The pulsars for which the outlier removal had the biggest effect were generally those with the smallest absolute RMs, indicating that weaker Faraday rotation signals are harder to constrain, as might be expected. For those pulsars with only one RM measurement, we quote the RM synthesis uncertainty on its measurement rather than the MAD. 

\begin{figure*}
\begin{tabular}{cc}
  \includegraphics[width=85mm]{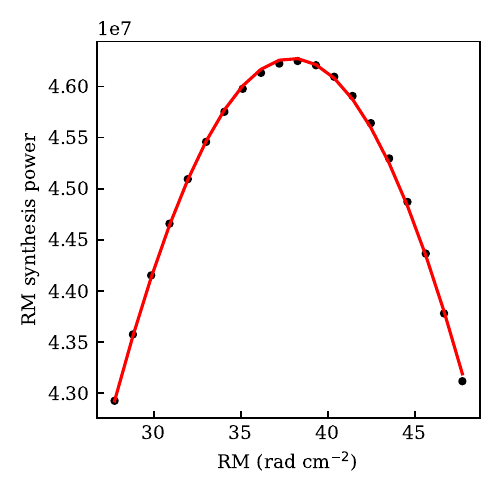} &   \includegraphics[width=85mm]{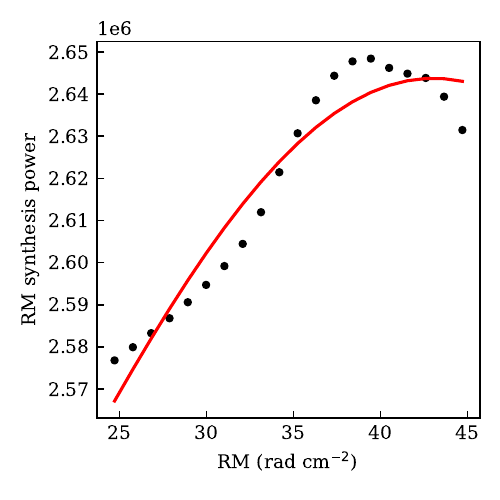} \\
 \includegraphics[width=85mm]{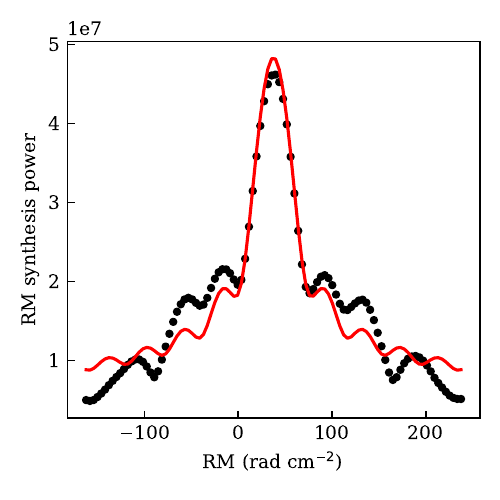} &   \includegraphics[width=85mm]{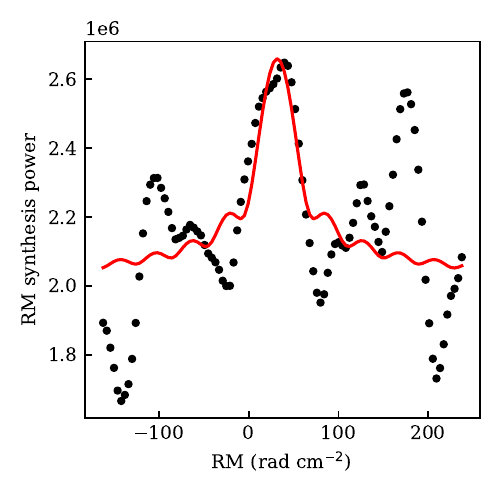} \\
\end{tabular}
\caption{Examples of the graphical output of the RM synthesis process for two observations of PSR~J0343$-$3000, taken at 13:37:46 on 16th May 2020 (top left and bottom left) and at 22:33:17 on 31st January 2022 (top right and bottom right). These two observations were taken as examples of a successful fit (left) and unsuccessful fit (right). The top two plots show the output when RM synthesis is run over a range of 40~rad~m$^{-2}$, from 28 to 48 rad~m$^{-2}$, and the bottom two plots show the same for a range of 400~rad~m$^{-2}$, from -160 to 240 rad~m$^{-2}$. The dotted line indicates the measured RM synthesis power at each trial RM, and the red line overlaid indicates the best fit sinc-like function, from which the best fit RM and its uncertainty are obtained.}
\label{fig:RMfitoutput}
\end{figure*}

We checked our metric for outlier removal by further categorising 722 RM measurement outputs by eye (639 ``good'' and 83 ``bad'') and using these to test the false positive and false negative rate. Fig. \ref{fig:RMfitoutput} shows the plotting output of the RM synthesis modelling, run twice over ranges of 20~rad~m$^{-2}$ and 400~rad~m$^{-2}$ (28~to~48 and $-$160~to~240~rad~m$^{-2}$ respectively in this case), for pulsar J0343$-$3000, showing both a successful RM fit from 16th May 2020 and a failed fit from 31st January 2022. The RM synthesis output should generate a distribution of power over trial RM that is representative of a sinc function, with a single defined maximum peak that corresponds to the best fit RM. For the failed measurement, it can be seen that there is no single clearly defined RM that corresponds to a power maximum. Comparing our outlier-removal technique to the by-eye assessment, we found a false positive rate of 4.8\% and a false negative rate of 4.5\% (true positive and true negative rates of 95.5\% and 95.2\% respectively). We are therefore confident that the outlier-removal technique, combined with the MAD, properly encompasses the failure rate of the RM modelling process.

\subsection{Dispersion measures, pulsar distances and electron density models}

We also require measurements of the pulsar DM and distance to estimate the magnetic field strength along the line of sight. For the DM measurements we use the census values published by \cite{Posselt2022y}. It is known that the choice of DM used to dedisperse a pulsar observation affects the best fit RM inferred \citep{Oswald2020}, and that pulsar DMs evolve with time \citep{Keith2024}. This means that for pulsars with multiple observations in our sample, using the same DM for every observation will introduce a small scatter on the inferred RM value. However, the magnitude of this scatter will be smaller than the dominant effect of the actual time-varying fluctuations in Faraday rotation strength, and will simply be encompassed as a small additional contribution to the MAD uncertainty we calculate for each RM. 

It is difficult to determine distances to pulsars accurately. Over 150 pulsars have independent distance measurements from parallaxes obtained through pulsar timing, optical techniques and very long baseline interferometry. The majority of these are millisecond pulsars \citep{Ocker2020} and for the rest of the data set the distances are currently undetermined. 
We therefore do not have all the information required in equation \ref{eq:Bpropto} to identify the distance to which our measurement of $\langle B_{\parallel}\rangle$ applies, and must therefore rely on a model of the distribution of the electron column density through the galaxy. Assuming this model, we can infer the distance to a given pulsar from its dispersion measure. We make use of the YMW16 model of Galactic electron density distribution for this purpose \citep{Yao2017}: for each pulsar we use the default distance given in the pulsar catalogue, which is either an independent distance measurement, or the inferred distance from the YMW16 model. Previous publications \citep[e.g.][]{Noutsos2008b} have used the NE2001 model instead \citep{Cordes2002}, however it was argued by \cite{Price2021a} that the YMW16 model is more accurate on average. They also demonstrated that both models have significant outliers and that the Galactic Halo should be incorporated into future models, factors which must be taken into account when judging the accuracy of any magnetic field modelling, however, we focus predominantly on the Galactic disk in this paper.

We note that the Galactic cartesian coordinates, $X$, $Y$ and $Z$, are definition-dependent, and furthermore are defined differently within {\sc psrcat} and {\sc psrqpy}\footnote{See details at \url{https://psrqpy.readthedocs.io/en/latest/index.html\#differences-with-the-atnf-pulsar-catalogue}, accessed 24th January 2024}. 
For clarity we therefore choose to calculate these coordinates ourselves from the pulsar distance, Galactic latitude $b$ and longitude $l$. We choose to align these measurements with the YMW16 model, since the majority of the distances we use are derived from this model. This places the origin such that the Sun is positioned at (0,8300,6)~pc, and the $X$ and $Y$ axes parallel to $l = 90\degree$ and $l = 180\degree$ respectively. We calculate the coordinates, in units of pc, as 
$X = r\sin(l)\cos(b)$, 
$Y = 8300 - r\cos(l)\cos(b)$, 
$Z = 6 + r\sin(b)$.

\subsection{Analytic magnetic field models}
\label{sec:analyticmodeldescription}

In 2008, \citeauthor{Noutsos2008b} investigated their sample of 150 pulsar RMs by comparing them to four analytic models of the Galactic magnetic field. They found that they did not have enough pulsar measurements to distinguish confidently between the different cases. We now perform a comparison of our modern data set with those same four models, to test whether the increased sample size of our data set improves the extent to which we are able to constrain the GMF analytically. Considerable time has passed since this last test of these models, and the field of GMF modelling has evolved considerably in the meantime. Nevertheless, we focus our attention on these four models specifically for the following reasons. First, a key way in which GMF modelling has evolved in recent years is to move away from prescriptive analytic models of these types, because of increasing evidence that they either do not provide enough information to encompass the complexity of the GMF, or because of insufficient capacity to constrain parameters using existing observations. Newer models are therefore less likely to be tractable to the type of simple analytic calculations described in the modelling process undertaken here. We argue that comparing our measurements to older analytic models retains important value. These models provide a representative range of types of spiral and non-spiral model, such that, even if the precise model parameters are inaccurate, we can  still test the relevance of symmetries/antisymmetries and contributing components to the disk field. In this light, the goal of model comparison becomes one of testing representative types of GMF structure, rather than constraining the specific model parameters, and we follow this up with a discussion of results in the context of more recent modelling approaches. Furthermore, by studying the same models as \cite{Noutsos2008b}, we can provide a direct comparison with previous pulsar-only modelling efforts, to understand the impact of increased numbers of measurements and make predictions about future advancements.

The four models in question are a simple dipolar-toroidal model (hence known as the DT model) and three more originally published by \cite{Tinyakov2002} (henceforth TT), \cite{Prouza2003} (PS) and \cite{Harari1999} (HMR). The last three of these are all variations of a logarithmic spiral disc field, with different pitch angles for the spirals, differing vertical suppressions of $\langle B_{\parallel}\rangle$ away from the Galactic plane, and differing symmetry/antisymmetry with respect to the Galactic plane. The PS model also includes dipolar and toroidal contributions in addition to the logarithmic spiral. Full descriptions of the models are given in \cite{Noutsos2008b} and the last three of these were originally collated and discussed by \cite{Kachelrieß2007}, however we also summarise their mathematical descriptions in Appendix \ref{appendix:ModelMaths} for ease of comparison.

In order to compare each model with the pulsar $\langle B_{\parallel}\rangle$ measurements, we must consider the impact on $\langle B_{\parallel}\rangle$ of averaging over both a changing electron column density and a changing field direction along a line of sight. To make this comparison, we wish to infer what we would detect observationally if the magnetic field followed a given analytical model. Since our observed pulsar parameters (the RM and DM) are dependent on the electron column density as well as the magnetic field, our model needs to include all of this information. We also need to integrate our models along the line of sight to make them comparable to observations. We therefore multiply the YMW16 model of electron column density and each of the four analytic models of the magnetic field in turn, and integrate this along the line of sight from the pulsar's position to Earth, to achieve a model RM. This integral requires knowledge of the pulsar's distance from Earth. If we have an independent distance measurement for a given pulsar, we use this in the integration. We also convert the independent distance into the inferred DM from the YMW16 model, for comparison. If the pulsar's independent distance is unknown, we use the YWM16 model to convert its DM to a model distance, and integrate over this to calculate the model RM and DM. We use {\sc pygedm} to read in values of electron column density from the YMW16 model, and to convert between DM and distance values \citep{Price2021a}. 

To obtain our integrated model RM and DM, we approximate the integrals as numerical summations over YMW16 electron column density and analytic magnetic field model. We sum along the line of sight from each of the modelled pulsar positions to the location of the Sun to obtain the modelled RM and DM:
\begin{equation}
    \textrm{RM}_{model} = C\sum_{\bm{r} = \bm{r_{p}}}^{\bm{r_{s}}}{n_{e}(\bm{r}) \bm{B}(\bm{r})\cdot\bm{\Delta L}}
\end{equation}
\begin{equation}
    \textrm{DM}_{model} = \sum_{\bm{r} = \bm{r_{p}}}^{\bm{r_{s}}}{n_{e}(\bm{r}) \lvert \bm{\Delta L}\rvert}
\end{equation}
where $\bm{r_{s}}$ and $\bm{r_{p}}$ are the positions of the Sun and the pulsar respectively, $C\approx 1.232$~rad~m$^{-2}$cm$^{3} \upmu$G$^{-1}$pc$^{-1}$ (see equations \ref{eq:RM} and \ref{eq:aveB} for its exact value), $\bm{\Delta L} = (\bm{r_{s}} - \bm{r_{p}})/N$, and $N$ is the number of steps of the summation, which we take to be 10,000. For a pulsar at a distance of a few kpc this gives a model resolution of a fraction of a parsec. The resolution will be slightly worse for pulsars at greater distances when using a fixed number of summation steps in this way, but we find that for the level of model accuracy we have available, the resolution given by $N=10,000$ is sufficient: all but one of the model DMs in the plane lie within 0.2~cm$^{-3}$pc of the value from the YMW16 model.

We note again the limitations of using the YMW16 model to infer pulsar distances. To investigate this, we compare the differences in values between the DM inferred by the YMW16 model and that observed, for pulsars with known distances. We find that although half (155 out of 307) lie within 20~cm$^{-3}$pc of each other, there exist outliers with differences of hundreds of cm$^{-3}$pc. Calculating the standard deviation, $\sigma$, of the differences between modelled and observed DM values for these pulsars, we find $\sigma = 111$~cm$^{-3}$pc for pulsars in the region $\lvert Z\rvert < 1$~kpc, and $\sigma = 36$~cm$^{-3}$pc for pulsars outside this region. This means that the vast majority of model discrepancies are for pulsars lying in the Galactic plane. By extension, it is likely that the model distances for many of the pulsars for which we do not have an independent distance are quite inaccurate: this should always be kept in mind.

\section{Results: measurements and their distribution on the sky}
\label{sec:results}

\subsection{Rotation measures, dispersion measures and distances}

\begin{table*}
    \caption{Table of time-averaged RM measurements for the Thousand-Pulsar-Array data set (first few lines here and the full table is given in the Supplementary Information). The columns indicate the name of the pulsar, the number of observations used in the analysis, the number of observations for which a successful RM measurement was obtained and used, the average RM inferred and the type of uncertainty used on that measurement. Where there is only one RM measurement, the uncertainty is that inferred from the RM synthesis technique; where there is more than one measurement, the median absolute deviation (MAD) is used. Pulsars missing information about the uncertainties of either the RM or DM (or both) are marked with an asterisk. The full table is available in the Supplementary Material online.}
    \begin{tabular}{lrrll}
        \hline
        PSRJ       &    $N_{obs}$ & $N_{obs}$               &                 RM    & Type of error \\
                   & (pre filter) & (post filter)           &(rad~m$^{-2}$)&               \\
        \hline
         J0034$-$0721 &                     10 &                      10 &        8.6 $\pm$ 0.3 &           MAD \\
         J0038$-$2501 &                      1 &                       1 &            9 $\pm$ 2 &       RMsynth \\
         J0045$-$7042 &                      2 &                       2 &       32.4 $\pm$ 0.4 &           MAD \\
         J0045$-$7319* &                     27 &                       5 &      $-$21.9 $\pm$ 0.4 &           MAD \\
         J0108$-$1431 &                     10 &                      10 &        3.5 $\pm$ 0.1 &           MAD \\
         J0113$-$7220 &                      2 &                       1 &          127 $\pm$ 2 &       RMsynth \\
         J0131$-$7310 &                      3 &                       1 &          $-$55 $\pm$ 5 &       RMsynth \\
         J0134$-$2937 &                     10 &                      10 &     15.93 $\pm$ 0.09 &           MAD \\
        ...\\
    \end{tabular}
    \label{tab:RMvaluestablesubset}
\end{table*}

\begin{table*}
    \caption{Table of RM, DM and distance measurements used for the magnetic field modelling, plus the inferred average magnetic field along the line of sight (brief subset presented here, full table in Supplementary Information). The ``Source'' column details whether the pulsar was taken from the pulsar catalogue (psrcat) or is part of the Thousand-Pulsar-Array data set (TPA). The ``Type of distance'' column indicates whether the distance used is independently measured or inferred from the DM using the YMW16 model. Pulsars missing information about the uncertainties of either the RM or DM (or both) are marked with an asterisk. The full table is available in the Supplementary Material online.}
    \begin{tabular}{lllllll}
        \hline
        PSRJ & Source & RM (rad~m$^{-2}$) &       DM (cm$^{-3}$pc) & Distance (kpc) & Type of distance & $\langle B_{\parallel}\rangle$ ($\upmu$G) \\
        \hline
        ...\\
        J0034$-$0534 & psrcat &                 $-$38 $\pm$ 17 &   13.76517 $\pm$ 0.00004 &          1.348 &            model &                                  $-$3 $\pm$ 2 \\
        J0034$-$0721 &    TPA &                8.6 $\pm$ 0.3 &           14.2 $\pm$ 0.2 &           1.03 &      independent &                             0.74 $\pm$ 0.02 \\
        J0036$-$1033 & psrcat &               $-$8.1 $\pm$ 0.7 &           23.1 $\pm$ 0.2 &           25.0 &            model &                            $-$0.43 $\pm$ 0.04 \\
        J0038$-$2501 &    TPA &                    9 $\pm$ 2 &            6.1 $\pm$ 0.1 &          0.604 &            model &                               1.7 $\pm$ 0.4 \\
        J0040+5716 & psrcat &               15.3 $\pm$ 0.2 &       92.515 $\pm$ 0.003 &           2.42 &            model &                           0.204 $\pm$ 0.003 \\
        J0045$-$7042 &    TPA &               32.4 $\pm$ 0.4 &           71.0 $\pm$ 0.9 &           59.7 &      independent &                             0.56 $\pm$ 0.01 \\
        J0045$-$7319* &    TPA &              $-$21.9 $\pm$ 0.4 &                  105.4 &           59.7 &      independent &                                 $-$0.256188 \\
        J0048+3412 & psrcat &              $-$83.3 $\pm$ 0.1 &         39.92 $\pm$ 0.01 &          4.501 &            model &                          $-$2.570 $\pm$ 0.003 \\
        ...\\
    \end{tabular}
    \label{tab:RMDMdistBvaluestablesubset}
\end{table*}

Table \ref{tab:RMvaluestablesubset} presents the first few lines of a table of all the time-averaged RMs of the TPA pulsar data set, for which an RM census was originally published by \cite{Posselt2022y}. These RMs are intended as more stable average RMs for the purpose of magnetic field modelling, thus where possible they are averages of several observations, as detailed in Section \ref{sec:RMmethods}. The full table of measurements is given in the Supplementary Material. 

Table \ref{tab:RMDMdistBvaluestablesubset} presents a representative subset of a table of RMs, DMs and distances used for the magnetic field modelling in this paper, including those taken from psrcat, and the inferred average magnetic field along the line of sight, $\langle B_{\parallel}\rangle$. In total the data set comprises the 741 pulsar RM and DM measurements taken from psrcat for the pulsars not part of the TPA data set, combined with measurements for the 1,097 pulsars presented in this publication. For four of the TPA pulsars (J1130$-$6807, J1226$-$3223, J1629$-$3825 and J1651$-$7642) we have an RM measurement but no published census DM: for these four we use the DM measurements taken from psrcat. For cases where we do not have an uncertainty estimate on either the RM, DM, or both, we do not quote an uncertainty for the inferred average magnetic field, and we mark the pulsar name with an asterisk. Again, we present the full table in the Supplementary Material.

We also present the RM measurements in the RMTable format developed by \citep{VanEck2023}. This format was developed to be a universal and flexible catalogue format for RM measurements, for use in galactic magnetism research, with regularly updated versions of the catalogue being maintained at 10.5281/zenodo.6702842. We have therefore reformatted all the RM measurements presented in Table \ref{tab:RMvaluestablesubset} into the RMTable format to enable easy integration into the catalogue. This too can be found in the Supplementary Material.

There are 39 pulsars in the Van Eck catalogue already, 36 of which also appear in this paper, and three of which do not. Those three have been obtained using RM synthesis of imaging observations in the GLEAM survey \citep{Riseley2020}. They do not have recorded RMs in {\sc psrcat}, nor do they form part of the sample of the Thousand-Pulsar-Array. For the other 36, we compare the two sets of RM measurements and find that 7 of the 36 pulsars have absolute RM differences greater than 15~rad~m$^{-2}$ across the two catalogues, with the largest being a difference of 630~rad~m$^{-2}$ for PSR~J1925$+$1720, with measurements of 444~rad~m$^{-2}$ in the Thousand-Pulsar-Array survey (published here), 437~rad~m$^{-2}$ in a previously published measurement given in {\sc psrcat}, and $-$186~rad~m$^{-2}$ in the GLEAM survey. Such a large difference seems best attributable to a measurement error, but on the whole, the smaller differences between measurements may partly be attributable to the time-fluctuations discussed above, and partly to the differences in observation method.

\subsection{Average magnetic field measurements in Galactic coordinates}
\label{sec:Bfieldonsky}

\begin{figure*}
    \centering
    \includegraphics[width=\textwidth]{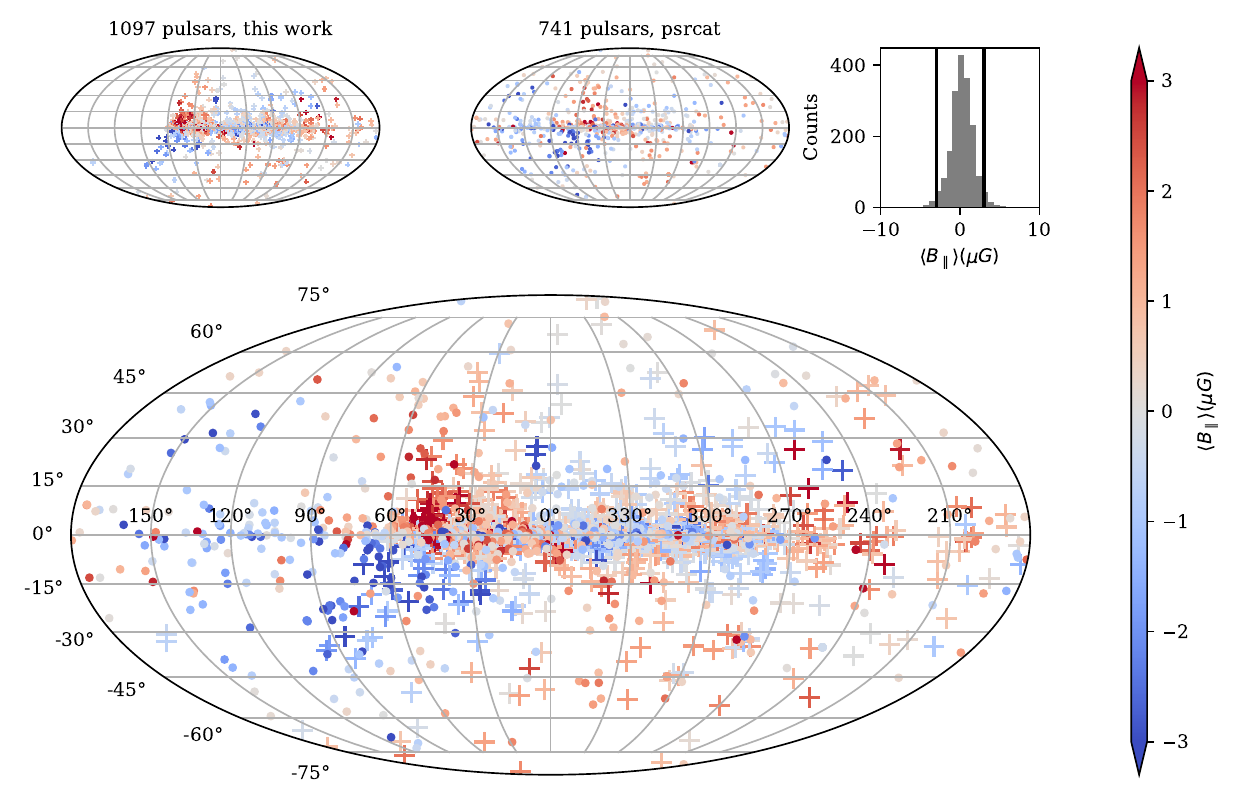}
    \caption{Main figure: sky map distribution of pulsars displayed as a function of Galactic latitude and longitude, where the colour indicates the average magnetic field component pointing along the line of sight, $\langle B_{\parallel}\rangle$. The magnetic field strength is capped at $\pm3{\upmu}$G. The crosses and dots indicate the two data sets used, which are also shown separately in the two subplots above. Top left (crosses): the subset of pulsar measurements taken in this work, from the Thousand-Pulsar-Array project. Top middle (dots): pre-existing pulsar measurements taken from the pulsar catalogue. Top right: histogram of $\langle B_{\parallel}\rangle$ measurements, with the x-axis range capped at $\pm10~{\upmu}$G. Two vertical lines mark $\pm3~{\upmu}$G.}
    \label{fig:psrBfieldsonsky}
\end{figure*}

Fig. \ref{fig:psrBfieldsonsky} shows the distribution of measurements of the average magnetic field component pointing along the line of sight to pulsars, $\langle B_{\parallel}\rangle$, as a function of Galactic latitude ($b$) and longitude ($l$). We present $\langle B_{\parallel}\rangle$ for 1,097 pulsars in the TPA data set and supplement this with a further 741 pulsars from the pulsar catalogue - these are particularly useful for the pulsars only observable in the Northern hemisphere. The colour of each point represents the magnitude of $\langle B_{\parallel}\rangle$, capped at $\pm~3~\upmu$G. Red points indicate that the sense of the observed magnetic field component points towards the Sun, blue indicates that it points away from the Sun. The overall view on the sky continues to support the broad schematic representation of a quadrupole configuration of the magnetic field that is antisymmetric with respect to both the Galactic plane ($b~=~0\degree$) and the line of Galactic longitude $l~=~8\degree$ \citep{Athanasiadis2004}. Our data set provides particularly good coverage of the Galactic plane, since most of the pulsar population are found there, and so we devote the majority of our analysis to modelling the configuration of the GMF within the plane.

Our map of magnetic field measurements to pulsars across the sky is qualitatively similar to the results of \cite{Hutschenreuter2023}, shown in their fig. 5, where they model the full average magnetic field across the sky using extragalactic radio sources (EGRS). Although their work maps out the average magnetic field across the full line of sight through the galaxy whereas our measurements show only the contribution up to the distance to a given pulsar, it is unsurprising that these two maps show many of the same features: this is evidence that the dominant contribution of the magnetic field to many lines of sight lies between the pulsar and the Earth, and is therefore sampled by both maps. We discuss the comparison to EGRS in greater detail in section \ref{sec:EGRS}.

\section{Investigating the Galactic plane}
\label{sec:galplaneresults}

\begin{figure*}
    \centering
    \includegraphics[height=0.9\textheight]{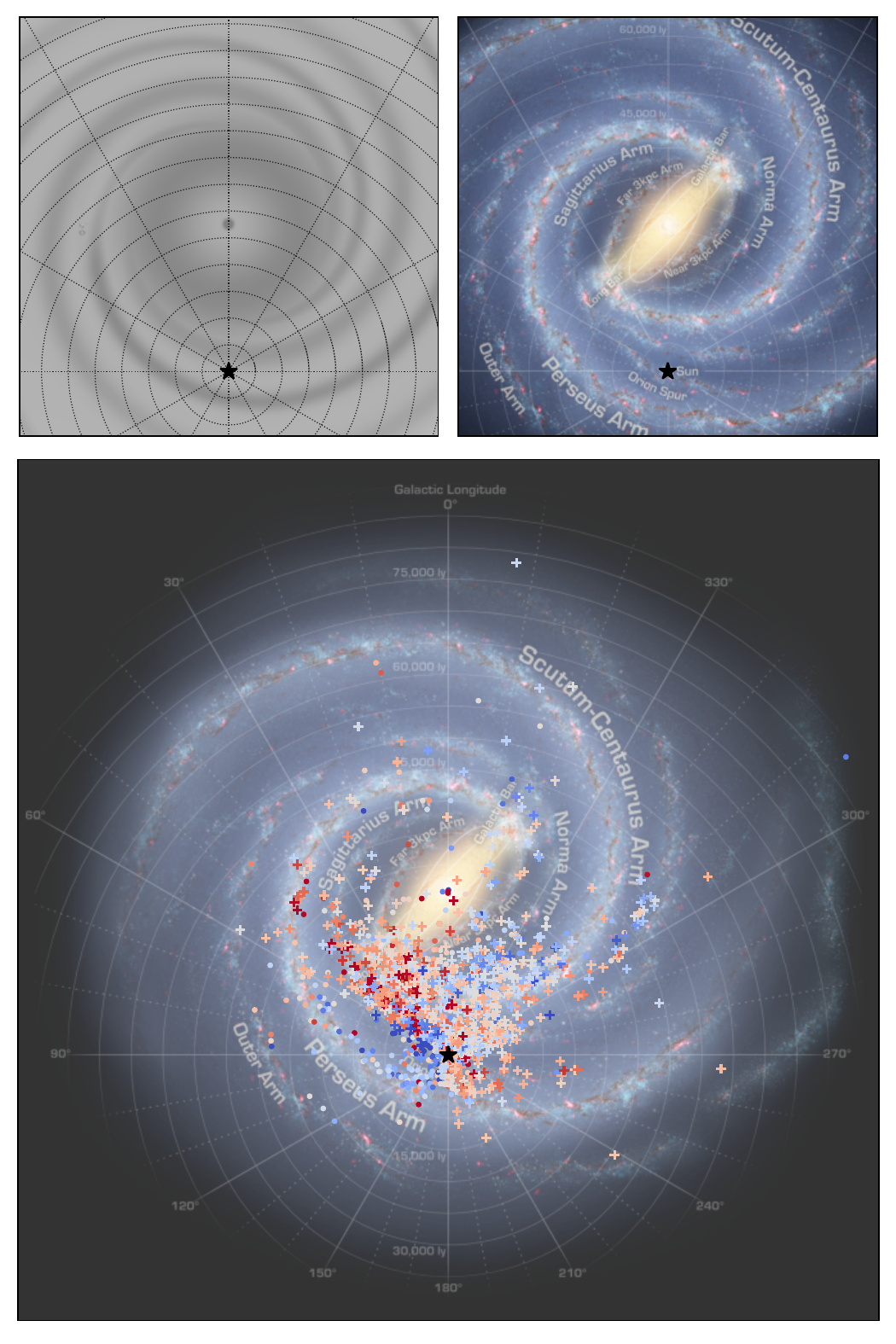}
    \caption{Top left: map of the YMW16 model of electron column density in the Milky Way. The overlaid dotted lines mark Galactic longitude increments of $30\degree$, and the circles mark increments of 5,000 lyr or 1,533 pc. Top right: labelled visual representation of the spiral arm structure of the Milky Way \citep{Churchwell2009}, sourced from \url{https://science.nasa.gov/resource/the-milky-way-galaxy}. Bottom: bird's eye view of the locations of the pulsars in the Galactic plane, overlaid on the visualization of the Milky Way spiral arms. The pulsar colour scheme depicts the average Galactic magnetic field component parallel to, and averaged along, the line of sight, in the same way as in Fig. \ref{fig:psrBfieldsonsky}. The location of the Sun is marked on all three subplots with a black star.}
    \label{fig:Bfieldinplane}
\end{figure*}

Fig. \ref{fig:Bfieldinplane} shows a representation of $\langle B_{\parallel}\rangle$ measurements to the subset of the pulsars that lie in the Galactic plane. Many papers define this as being within $\pm 8\degree$ of Galactic latitude from the plane, after \cite{Han1999}. We instead choose to define this as being within 1~kpc of the Galactic plane in Galactic Cartesian coordinates, i.e. $\lvert Z\rvert < 1$~kpc. This definition relies on the YMW16 model to calculate $Z$, but, since our investigations rely on this model anyway, this has the advantage of giving a self-consistent definition of Galactic plane pulsars within our modelling set-up. Our TPA data set are marked with crosses and the pulsar catalogue measurements are marked with circles, as in Fig. \ref{fig:psrBfieldsonsky}. The location of the Sun is marked with a black star, and dotted lines mark lines of Galactic longitude, as labelled. We show two representations of the large-scale structures of the galaxy in inset figures: the YMW16 model of electron distribution and an artist's impression of Galactic structure displaying the names of the Galactic spiral arms\footnote{Sourced from: \url{https://science.nasa.gov/resource/the-milky-way-galaxy}, accessed 1st December 2023} \citep{Churchwell2009}. The average magnetic field measurements to pulsars are superimposed onto the image of Galactic structure, to compare the distribution of the pulsars with the positions of spiral arms. 

Most pulsars are located upon the spiral arms of the YMW16 model. Although it is expected that there will be more pulsars located in spiral arms than between them, it should be noted that the reliance upon the YMW16 model for the placing of individual pulsars means that care should be taken about not relying too heavily on the modelled pulsar positions when making inferences about the magnetic field structure. From Fig. \ref{fig:Bfieldinplane} we can see that, although the YMW16 spiral arms can be identified as corresponding to the named stellar spiral arms, the winding of the two sets of spiral arms does not always align. A good example is the looser curvature of the Perseus arm in the region $90\degree < l < 270\degree$ compared to the tighter curvature of the YMW16 spiral arm in that longitude region. This means that the pulsars in this region are lie upon the YMW16 spiral arm and therefore appear to lie between the Perseus and Scutum-Centaurus arms. Without independent distance measurements this positional uncertainty is difficult to break. This will have an impact on the extent to which we can infer whether the GMF follows the spiral arms or has a different pitch angle.

As discussed in section \ref{sec:Bfieldonsky}, the GMF appears antisymmetric with respect to the Galactic plane. From this point onwards we therefore split our dataset into pulsars sitting just above and just below the Galactic plane, to assess the extent to which this antisymmetry impacts our modelling results.

\subsection{Simple tests of a spiral configuration}
\label{sec:spiraltests}

\begin{figure*}
    \centering
    \includegraphics[height=0.85\textheight]{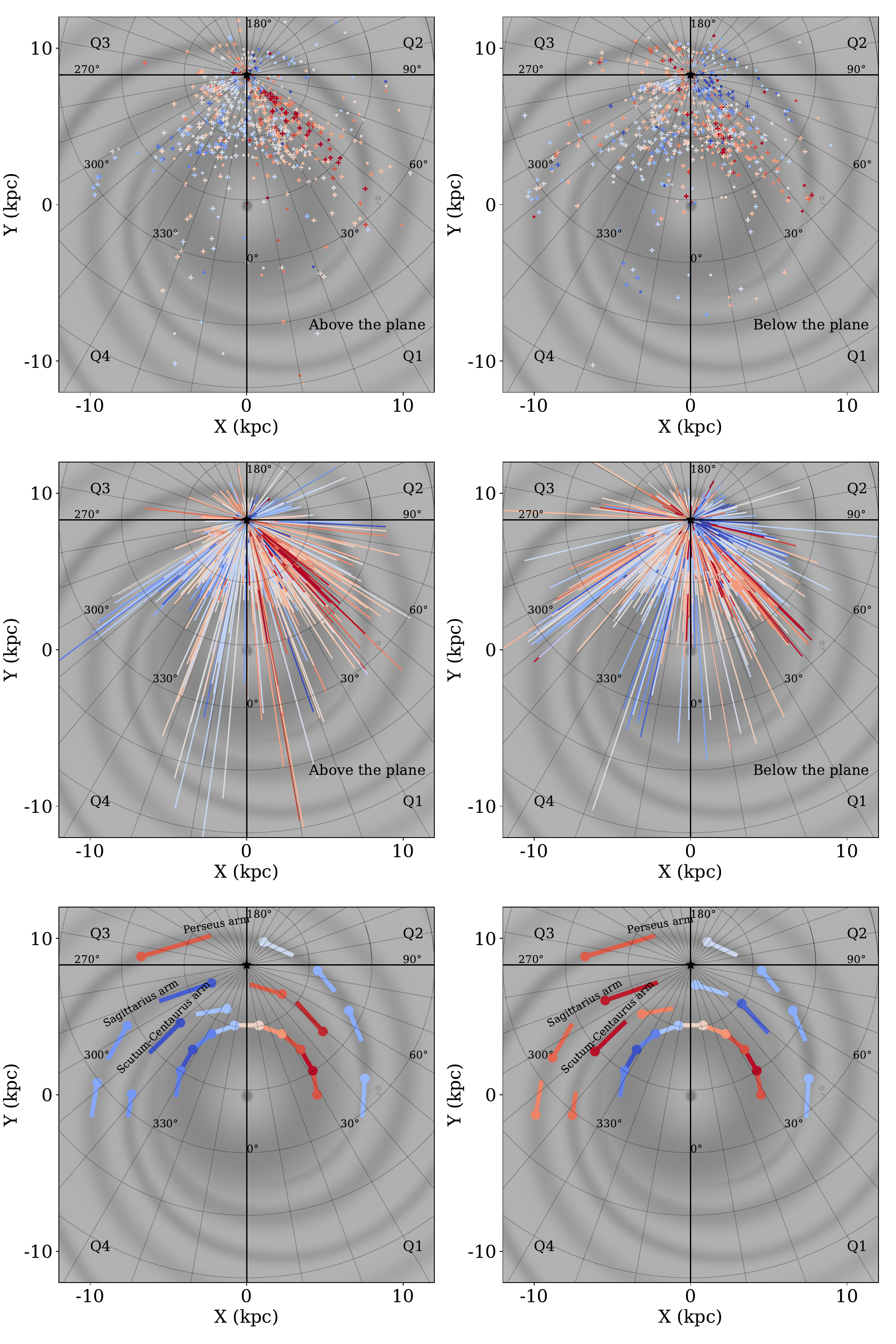}
    \caption{Top left: the Galactic disk pulsars sitting above the Galactic plane $Z = 0$ ($0 \leq \lvert Z\rvert < 1$~kpc). The colour scheme depicts the GMF component parallel to and averaged along the line of sight to the pulsar, as in Fig. \ref{fig:Bfieldinplane}. Top right: the equivalent below the Galactic plane ($-1 < \lvert Z\rvert \leq 0$~kpc). Middle left and middle right: As for top left and top right, but now with the measurements of $\langle B_{\parallel}\rangle$ represented as lines rather than points, to emphasize the fact that these measurements are averages along the whole line of sight. Bottom left and bottom right: visualizations of a simple spiral model of the Galactic magnetic field above (left) and below (right) the Galactic plane, indicated by coloured vectors. The direction of each vector points from the circular end towards the squared end. The length of each vector has no significance and is selected only to guide the eye in comparing the direction of the vector with the curvature of the spiral arm. The colour of each vector indicates the angle $\theta$ that the vector makes with the line of sight to the Sun, calculated from the dot product. In each figure the YMW16 electron column density model is plotted in the background, the location of the Sun is marked with a black star and increments of 10$\degree$ of Galactic longitude $l$ are indicated with thin radial lines, with labels given in increments of $30\degree$. Circles are plotted to mark increments of 4~kpc radial distance from the Sun. The line $l = 0\degree$ is marked with a thick black line going vertically downwards from the Sun. The lines $l = 90\degree$, $180\degree$ and $270\degree$ are similarly marked with thick black lines. Further details are given in the text.}
    \label{fig:simplespiralmodel}
\end{figure*}

The axisymmetric and bisymmetric spiral models of the GMF propose that the magnetic field follows the structure of the Galactic spiral arms. In the axisymmetric case, the magnetic field would point along the clockwise direction of all of the arms; in the bisymmetric case the field flows inwards from one end of a spiral and outwards on the spiral on the opposite side of the centre. That could, for example, correspond to the field flowing inwards along the Perseus arm and outwards along the Scutum-Centaurus arm, and similarly inwards along the Norma/Outer arm and outwards along the Sagittarius arm, or the equivalent but all flowing in the opposite direction.

In Fig. \ref{fig:simplespiralmodel}, we create a simple visualisation of some of the observable effects of a spiral model. The upper four subplots of the figure show the distribution of $\langle B_{\parallel}\rangle$ within the plane again, as in Fig. \ref{fig:Bfieldinplane}, now split into the upper/lower halves of the Galactic plane and superimposed against the YMW16 electron density model, since this is what is used to identify the locations of pulsars. The X-Y components of the Galactic coordinate system label the axes, the four Galactic quadrants are labelled Q1 to Q4 and dotted lines indicate Galactic longitudes in increments of $l = 10\degree$, starting from $0\degree$ along the line $X = 0$, $Y < 0$ that passes through the Sun and the Galactic centre. It should be stressed that our observations show the average magnetic field component pointing along the line of sight, and not the magnetic field measurement at the position of pulsar in question. For this reason, we have plotted the figures twice, once with dots/crosses marking the positions of pulsars, and once with lines extending from the Sun along the lines of sight to those positions.

The lowest pair of subplots in the figure shows a series of vectors pointing along the directions of the spiral arms of the YMW16 model. We chose the direction of the magnetic field along each spiral arm to match roughly with the observed directions of the average magnetic field to the pulsars. We associate the YMW16 spiral arms with named Galactic spiral arms as follows: moving from top to bottom on the figure the arms with vectors overlaid are the Perseus, Sagittarius and Scutum-Centaurus arms (labelled respectively on the figure). The innermost region of the YMW16 model has a much-simplified circular symmetry and so our associated semi-circle of vectors roughly encompasses regions of the Norma arm, Near-3kpc arm and Galactic bar. To compare with the pulsars above the plane ($0 \leq \lvert Z\rvert < 1$~kpc, bottom left plot) we directed the field along the Perseus arm inwards, Sagittarius arm outwards, Scutum-Centaurus arm outwards, inner region anticlockwise circle. To compare with the pulsars below the plane ($-1 < \lvert Z\rvert \leq 0$~kpc, bottom right plot) we directed the Perseus arm field inwards (symmetric about the plane), Sagittarius arm inwards (anti-symmetric), Scutum-Centaurus arm inwards (anti-symmetric), inner region anticlockwise circle (symmetric). Where vector colour is strongest, the vector direction is very close to the line of sight ($\theta \sim 0\degree$ for red or $\theta \sim 180\degree$ for blue), and we would expect to observe a strong measurement of $\langle B_{\parallel}\rangle$ from this region. Where the colour is pale, the magnetic field direction is closer to $90\degree$ away from the Sun, and we would expect to see little or no magnetic field from this region. 

It is important to note that this simple visualization does not include any information about the actual magnitude of $\langle B_{\parallel}\rangle$, which might be expected to be stronger closer to the Galactic centre, nor does it account for the fact that our actual observations depend on the average along the whole line of sight. For example, the vector closest to the bottom right of the bottom left plot in the figure ($l \sim 45\degree$, distance from Sun $\sim 12$~kpc) is a pale blue colour, however the line of sight to this region would pass through a region where the magnetic field points directly towards the Sun. This means that the average magnetic field inferred along the line of sight to this region would encompass everything along the line of sight, and the field would likely be observed to be red (positive) rather than blue (negative).

Comparing this simple visualization to the observations, it is clear that the assumption that the magnetic field points along the spiral arms is reasonably successful in explaining the overall magnetic field directions and, to some extent, the varying strengths of observed magnetic field, but only if the direction along different spiral arms is allowed to be different. This pushes the model more in favour of a bisymmetric spiral than an axisymmetric spiral: indeed the visualisation comparing to the pulsars above the Galactic plane follows exactly the bisymmetric prescription described at the start of this section. However, there same is not true for below the plane  ($-1 < \lvert Z\rvert \leq 0$~kpc), where only the Sagittarius and Scutum-Centaurus arms have their magnetic field direction reversed. 

This raises an interesting question about the anti-symmetry of the GMF with respect to the plane. It is well known that the field appears anti-symmetric in quadrants 1 and 4, but symmetric in quadrants 2 and 3. But our viewing location at the Sun is unlikely to be a special case (although we note that \cite{He2021} locate the Galaxy corotation radius close to the Solar circle), so the quadrant distinction ought to be arbitrary unless it is reflecting some limitation of our viewing geometry. This simple spiral visualisation model can account for that, by changing the symmetry/anti-symmetry split from referring to quadrants to referring to different spiral arms. If the Perseus spiral arm is symmetric with respect to the plane, but the Scutum-Centaurus and Sagittarius arms are not, then we will observe the majority of the symmetry/anti-symmetry split to fall along the quadrants as shown, since the Scutum-Centaurus and Sagittarius arms lie entirely in Q1 and Q4. The Perseus arm extends into Q1, but the magnetic field there is likely to be mostly masked by the stronger impact of the Sagittarius arm which lies along the line of sight, and into Q4, where it is sufficiently far from the Sun that we have only one pulsar lying in the arm, and again the average magnetic field along this line of sight will also be dominated by the Sagittarius arm. We note also that the model of the Sagittarius arm being antisymmetric with respect to the plane was independently proposed by \cite{Ma2020}, which supports this description in that region. The cause of this change in symmetry for different spiral arms could be that the magnetic field of the Perseus arm is inclined with respect to the Galactic plane. If the stellar/gas density follows a similar inclination to the magnetic field, it is possible that stellar/gas tracers could be employed to test this, but further investigation along these lines is beyond the scope of this work.

Another interesting question to address is whether a ring-shaped magnetic field is a more useful or relevant picture than a spiral. In this simple visualization we make use of a ring-shaped field closest to the Galactic centre: visually this seems a reasonable approximation to the observations, but the increased complexity of structures closer to the Galactic centre makes the picture more difficult to ascertain. The reality of the Milky Way having a bar structure in the centre may affect the magnetic field in two ways: first it may directly alter the structure of the field, and secondly the electron density distribution may be altered from the circular region given in the YMW16 model, affecting the inferred distances to pulsars. There is considerable discussion in the literature about an annular region in which the field direction opposes the rest of the field. This may also be accounted for by our description here of the Scutum-Centaurus and Sagittarius arm fields changing direction below the plane  ($-1 < \lvert Z\rvert \leq 0$~kpc), whereas the inner field region does not. We discuss this more in section \ref{sec:annular}. 

There remains one key region that does not conform to the prevailing spiral arm direction suggested by our simple model: the region at $280\degree < l < 300\degree$ within $\sim 3$~kpc of the Sun. In this region the prevailing magnetic field direction is opposite to that expected from the anti-clockwise and clockwise field directions inferred above and below the plane respectively. We return to discussion of this region in the context of EGRS RM measurements in section \ref{sec:EGRS}.

\subsection{Comparison with extragalactic radio sources}
\label{sec:EGRS}
\begin{figure*}
    \centering
    \includegraphics[width=\textwidth]{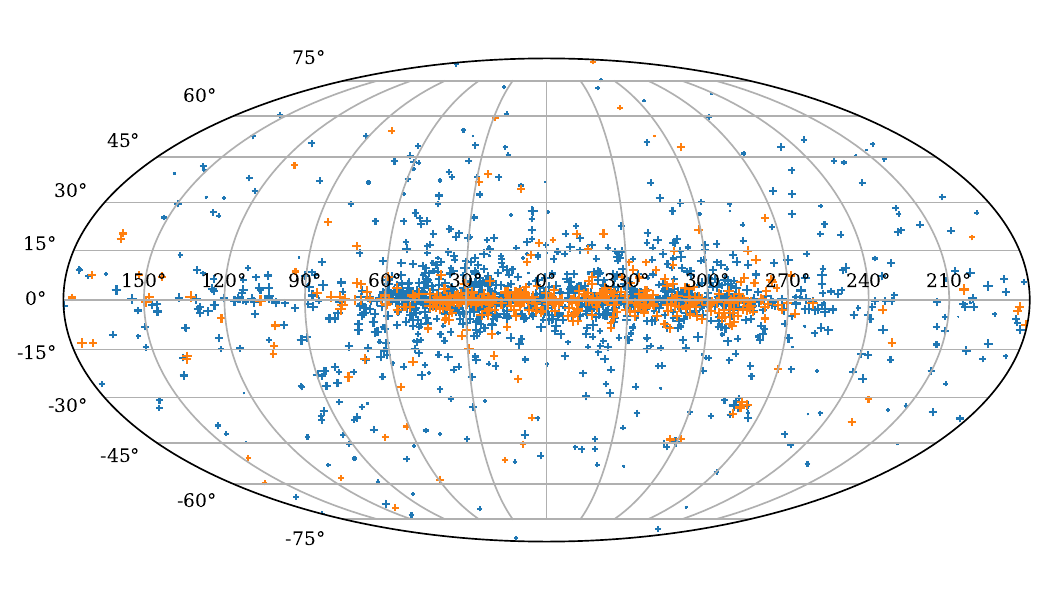}
    \caption{Sky map distribution of pulsars as a function of Galactic latitude and longitude, coloured for comparison with RM measurements to EGRS along the same lines of sight. Blue crosses indicate that pulsar and EGRS RMs have the same signs, whereas orange crosses indicate different signs. The size of cross is scaled as $10\log_{10}(\lvert RM_{EGRS}~-~RM_{pulsar}\rvert)$ to represent the magnitude of difference of EGRS and pulsar RMs. Further details are given in the text.}
    \label{fig:EGRSpsrOnsky}
\end{figure*}

We compare the RMs of our pulsar data set with the RMs of EGRS positioned directly behind those pulsars as viewed from Earth. For this we use the Faraday rotation sky map constructed by \cite{Hutschenreuter2020}\footnote{Source: \url{https://wwwmpa.mpa-garching.mpg.de/~ensslin/research/data/faraday2020.html}, accessed 1st March 2023.}. We identify the relevant EGRS RM per pulsar by converting the Galactic latitude and longitude of the pulsar into the appropriate pixel in the HEALPix\footnote{\url{https://healpix.jpl.nasa.gov/}} scheme (resolution Nside=512), and using the RM associated with that pixel. Note that we are therefore not comparing pulsar RMs with the RMs of specific radio sources, but with the Faraday rotation value inferred at that point on the reconstructed sky map, which has a resolution of $46.8~\textrm{arcmin}^{2}$. 

Fig. \ref{fig:EGRSpsrOnsky} compares the pulsar RMs with the EGRS RMs positioned along the same line of sight. Crosses mark the positions on the sky in terms of Galactic latitude and longitude, and the size of the cross indicates the magnitude of the difference of the two RMs, scaled to a reasonable visual comparison as $10\log_{10}(\lvert RM_{EGRS}~-~RM_{pulsar}\rvert)$, so that larger crosses indicate cases where the RM difference is larger. The colour of the cross indicates whether the RM of the pulsar and the EGRS have the same sign (blue) or different signs (orange). If the latter, then there must be at least one reversal of the magnetic field with a strong effect somewhere along the line of sight between the pulsar and the EGRS. In cases where the pulsar and EGRS RM signs are different, but the absolute difference is less than $3\sigma$, where $\sigma$ is defined as the error on the RM measurement, we conclude that the fact that their signs are different from each other is insignificant. We therefore remove these points from the figure. 

It can be seen that, away from the Galactic plane, most of the crosses are blue, i.e. the pulsar and the EGRS RMs have the same sign. It can also be seen that, as you move away from the Galactic plane, the size of the crosses gets smaller – the pulsar measurements and the EGRS measurements are more similar to each other. This suggests that, away from the plane, the field is weaker and more homogeneous, so there are likely to be fewer field reversals along the line of sight. Furthermore, the field is likely strongest closest to the middle of the galaxy, so in most cases the dominant contribution to the field along the whole line of sight is likely to be sampled by the pulsar as well as the EGRS, unless the pulsar happens to be very nearby. Out of the plane, it is therefore more useful to use EGRS maps to probe $\langle B_{\parallel}\rangle$, since the sky coverage is vastly improved and the view from within the galaxy itself does not provide much additional information.

The majority of the orange crosses in Fig. \ref{fig:EGRSpsrOnsky} lie along the Galactic plane, and the size of these crosses is larger, indicating a strong difference between pulsar and EGRS RMs along these lines of sight. This makes sense: in the plane the field is strongest and the effect of the spiral arms and more extreme variations in localised regions are likely to cause observable field reversals along the line of sight. We therefore investigate the comparison in the plane more closely.

\begin{figure*}
    \centering
    \includegraphics[width=\textwidth]{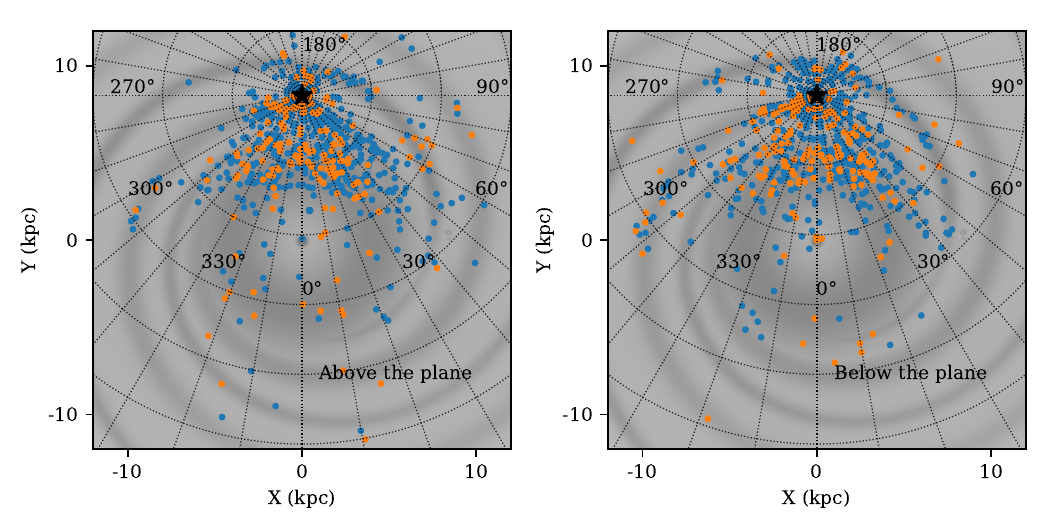}
    \caption{Bird's eye view of pulsars in the Galactic disk and above and below the Galactic plane ($0 \leq \lvert Z\rvert < 1$~kpc and $-1 < \lvert Z\rvert \leq 0$~kpc respectively), coloured to show the comparison of pulsar and EGRS RMs, where blue indicates the same sign and orange indicates opposite signs, as described for Fig. \ref{fig:EGRSpsrOnsky}.}
    \label{fig:EGRSpsrInplane}
\end{figure*}

In Fig. \ref{fig:EGRSpsrInplane} we compare the signs of the pulsar and EGRS RMs in the plane, splitting into views above and below the plane ($0 \leq \lvert Z\rvert < 1$~kpc and $-1 < \lvert Z\rvert \leq 0$~kpc respectively). We use this split view for two reasons, first for easier direct comparison with Fig. \ref{fig:simplespiralmodel}, and second as a useful check for identifying unusual regions in the galaxy. If the behaviour appears different above and below the plane, that implies the existence of a localised region that is dominating the apparent magnetic field either above or below the plane, because otherwise we would not necessarily expect to see an overall difference in the two plots. We begin by comparing the pulsar and EGRS RM signs in the plane as a whole, before discussing some particular regions of interest.

We consider what results we would expect to see from our simple spiral model and compare this to the distribution of similar (blue points) and different (orange points) RM signs between the pulsars and the EGRS. We see that overall the orange points are congregated both very close to the Sun and at larger distances from the Sun towards the Galactic centre. Following a line of sight through the Galactic centre, we would expect to see a magnetic field that is stronger and more variable than that in the halo. As a result, we expect RM measurements to locations beyond the Galactic centre to include contributions from areas where the field direction is reversed. This means that along lines of sight that pass close to the Galactic centre, roughly $-40\degree < l < 40\degree$, we expect a fairly even distribution of blue and orange points, respectively encompassing even and odd numbers of field reversals in the Galactic centre region between the pulsar and the EGRS: these expectations are fulfilled by the observations. By contrast, in regions of Fig. \ref{fig:simplespiralmodel} where we observe dark red or blue, that is, strong $\langle B_{\parallel}\rangle$, we expect that that region provides the magnetic field contribution that dominates what we see along that line of sight. This means we would expect same sign of RM from the EGRS - this is indeed the case along the region $40\degree < l < 60\degree$. 

When looking away from the Galactic centre and towards quadrants 2 and 3, the observed $\langle B_{\parallel}\rangle$ should be weaker, so we expect the majority of pulsars and EGRS to have the same sign, because the region closest to the Sun will have the strongest impact on the observed $\langle B_{\parallel}\rangle$ and this region is common to both lines of sight - this is again seen to be the case. 

It is particularly difficult to constrain our understanding of the magnetic field at large distances due to the limited number of pulsar observations. Future pulsar discoveries will fill out the map, but pulsar discoveries from the region beyond the Galactic centre will be particularly hampered by scattering of the radio emission by intervening material. Comparing pulsar and EGRS RMs indicates considerable complexity in the magnetic field viewed towards the Galactic centre, but at present we have insufficient data to determine the extent to which this is caused by sign reversals in the global field rather than by local regions being the dominant contribution to the average magnetic field.

On the whole, we find that the results from comparing the signs of the RMs of pulsars and EGRS follow our expectations from our simple spiral model as laid out above, however there are two regions that stand out for being different. This is particularly evident when we compare behaviour above and below the Galactic plane. The first is a region above the plane in the Perseus arm, around 8~kpc from the Sun and at $60\degree < l < 70\degree$, where there is a cluster of orange points not replicated below the plane ($-1 < \lvert Z\rvert \leq 0$~kpc). Comparing with Fig. \ref{fig:simplespiralmodel}, we see that we measure a magnetic field pointing towards the Sun in this region, which is different from what would be predicted due to a clockwise field in the Perseus arm. This suggests that the pulsars in this region are probing a special local field configuration of some sort, which is why the behaviour stands out against the EGRS background. The second region of interest is on the Sagittarius arm, close to the Sun and at $280\degree < l < 300\degree$. Here we see that the sign relative to the EGRS background changes with distance from the Sun, going from opposite (orange) to the same (blue) as you move away from the Sun. It is also slightly different above and below the plane. This region also stands out as not fitting with the expected spiral arm direction indicated by Fig. \ref{fig:simplespiralmodel}. This could suggest that there is a region slightly further from the Sun that locally has the field pointing in the opposite direction to that expected from the spiral arm direction: this region dominates the field direction both for pulsars from this region and for the EGRS beyond it. To test this properly we would need to discover more pulsars in the region $280\degree < l < 300\degree$ but at larger distances from the Sun, to further probe this relationship with distance.

\section{Testing analytic models}
\label{sec:analyticmodels}

\subsection{Comparison with four analytic models}
\label{subsec:comparemodels}

\begin{figure}
    \centering
    \includegraphics[width=\columnwidth]{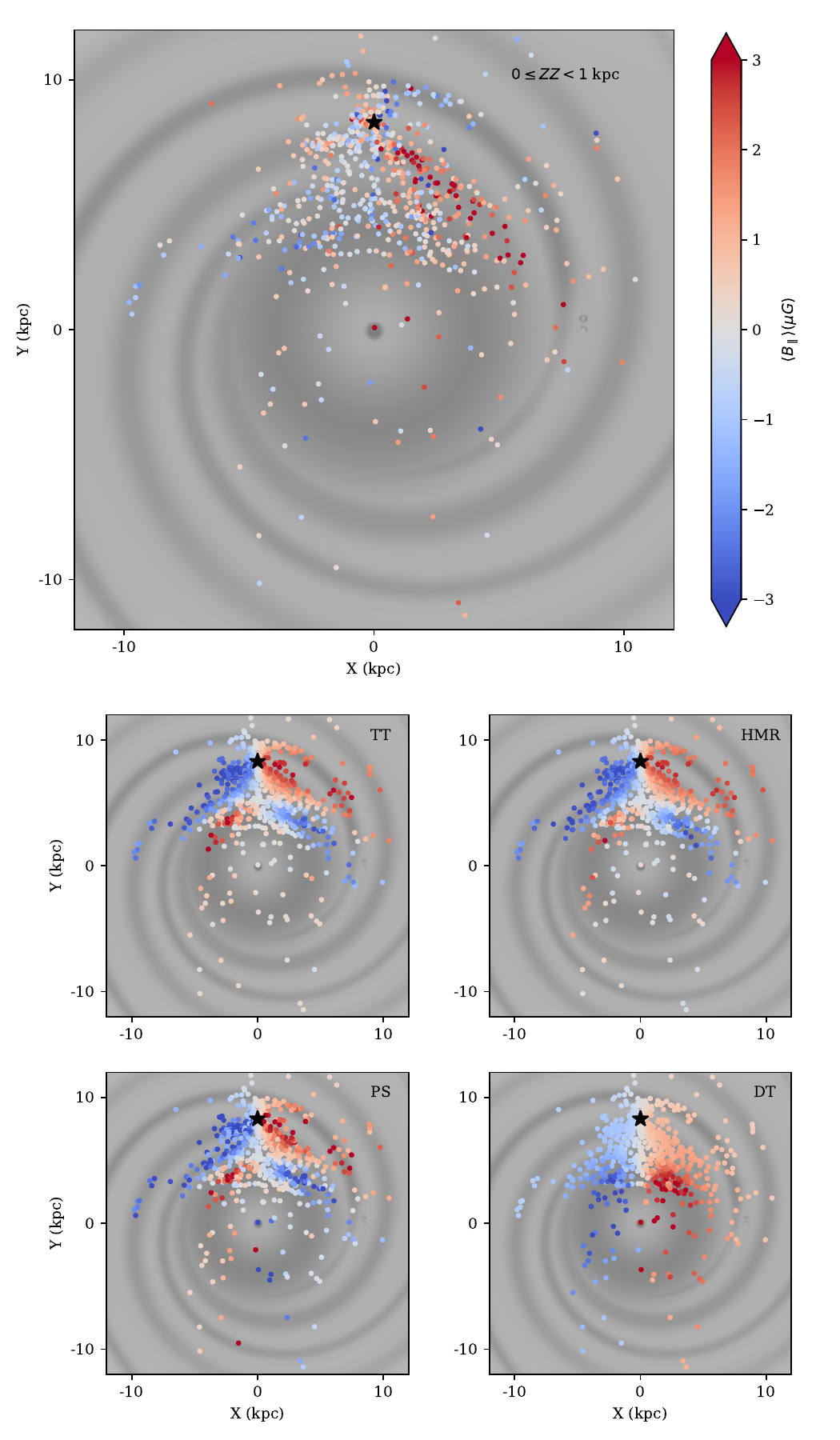}
    \caption{Top: pulsar magnetic fields in the top half of the Galactic disk ($0 \leq \lvert Z\rvert < 1$~kpc), distributed and coloured identically to the top left subplot of Fig. \ref{fig:simplespiralmodel}. Below: modelled $\langle B_{\parallel}\rangle$ for the same locations for four analytic models. All magnetic field strengths shown here are capped at $\pm3{\upmu}$G. Further details of the analytic models are given in the text. }
    \label{fig:fig7_modelsabove}
\end{figure}

\begin{figure}
    \centering
    \includegraphics[width=\columnwidth]{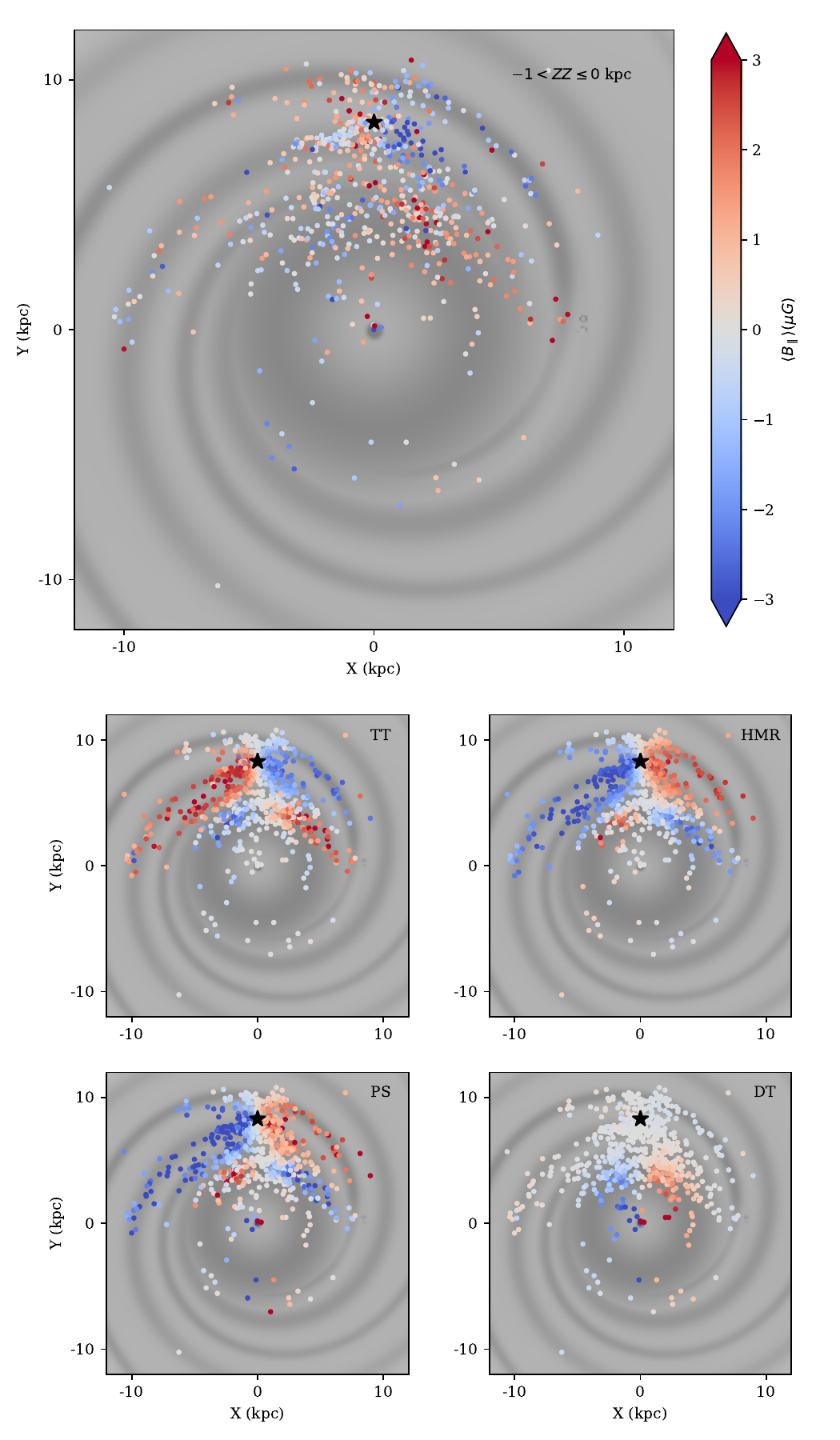}
    \caption{As for Fig. \ref{fig:fig7_modelsabove}, this figure depicts pulsar measurements of $\langle B_{\parallel}\rangle$ (top) and corresponding modelling results for four analytic models (below), but now for pulsars in the bottom half of the Galactic plane ($-1 < \lvert Z\rvert \leq 0$~kpc).}
    \label{fig:fig8_modelsbelow}
\end{figure}

We now investigate how the analytic magnetic field models, described in section \ref{sec:analyticmodeldescription}, compare to our observations. Figures \ref{fig:fig7_modelsabove} and \ref{fig:fig8_modelsbelow} show visual comparisons of how $\langle B_{\parallel}\rangle$ compares for the data and the four models, split to show the comparison above (Fig. \ref{fig:fig7_modelsabove}) and below (Fig. \ref{fig:fig8_modelsbelow}) the Galactic plane. Visually, it appears that of these four models the TT model best describes the data, because the HMR and PS models are symmetric and therefore do not represent the magnetic field direction well below the plane  ($-1 < \lvert Z\rvert \leq 0$~kpc), and the DT model contains no field reversals, unlike what can be seen in the observations.

\begin{figure}
    \centering
    \includegraphics[width=0.6\columnwidth]{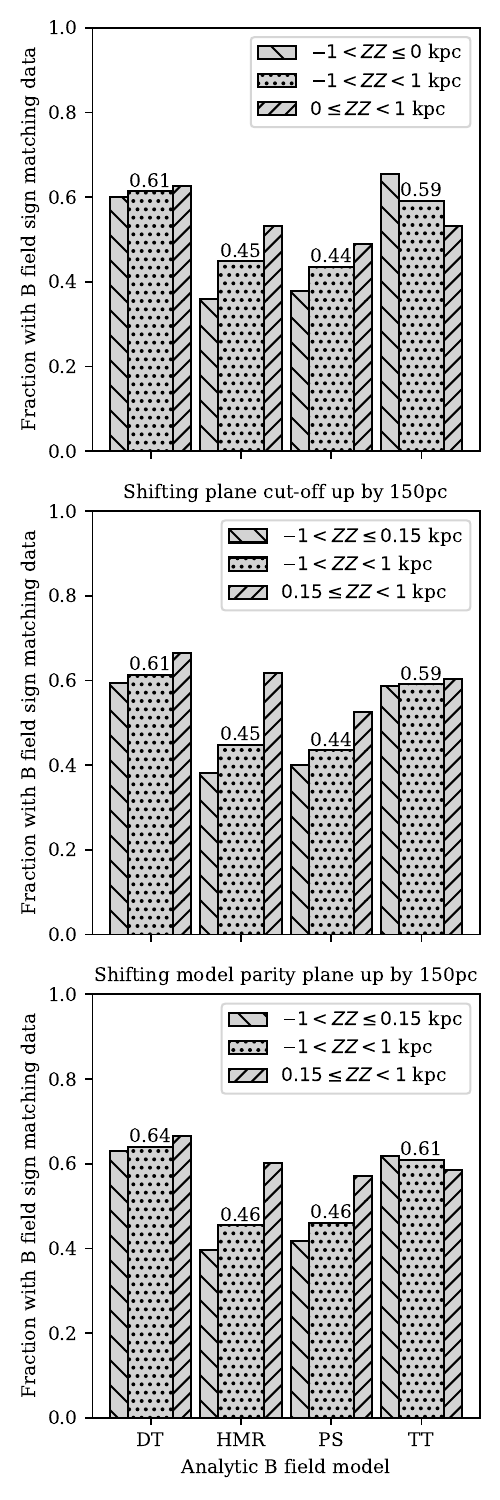}
    \caption{Top: bar chart showing fractions of pulsars for which the observations and the analytic model have $\langle B_{\parallel}\rangle$ measurements with the same sign. For each of the four models (details of them in the text), the bars show the total fraction of pulsars in the Galactic plane (dotted), the subset below $Z = 0$ (left diagonal) and the subset above $Z = 0$ (right diagonal). Middle: the same bar chart, but now the subsets are divided with respect to the plane $Z = 0.15$~kpc. Bottom: same bar chart, but now the zero-plane of the analytic models has been shifted up by $Z = 0.15$~kpc and the model outputs have been recalculated and then compared to the data set. }
    \label{fig:fig9_barchart}
\end{figure}

Next, we compare the models in a more quantitative way, by counting the number of measurements of $\langle B_{\parallel}\rangle$ in each model that have the same sign as the equivalent data measurements. We show the results of this as a bar chart in the top subplot of Fig. \ref{fig:fig9_barchart}, as fractions of the measurements with a matching magnetic field direction (sign) above the plane ($0 \leq \lvert Z\rvert < 1$~kpc), below the plane ($-1 < \lvert Z\rvert \leq 0$~kpc), and in total. We find that, despite its obvious systematic residuals, quantitatively the DT model performs best, with 62\% of the observations having a matching magnetic field direction. The TT model comes a close second, with a 59\% match. We note, however, that the three logarithmic spiral models show discrepancies in how well they perform above and below the plane, and so we investigate shifting the position of where we split the plane in two, to see if this improves these discrepancies. 

We investigate this in two ways. First, we simply shift the position of our cut from $Z = 0$~kpc to $Z = 0.15$~kpc, so that some of the pulsars being counted as ``above the plane'' are now identified as being ``below the plane''. We tested $Z = 0.1$, $0.15$ and $0.2$~kpc, all of which are shifts of no more than 10\% of the thickness of our defined Galactic plane ($\lvert Z\rvert < 1$~kpc). We find, as shown in the middle subplot of Fig. \ref{fig:fig9_barchart}, that shifting the cut evens out the fractions of pulsars for which the magnetic field direction matches the data for the TT model, and worsens the discrepancy for the other three models. This is most noticeably true for $Z = 0.15$~kpc. This suggests, but does not confirm, that the pulsars would be better modelled by an antisymmetric magnetic field which is shifted so that its plane of parity is slightly above the zero point of the stellar plane. We do not attempt a more careful fit to find the best shift of $Z$, because the goal here is to perform an indicative test rather than to place undue importance on a single variable. We then rerun all of the model calculations, but with this shift of $Z = 0.15$~kpc included in the model itself, and then repeat the calculation of agreement of magnetic field signs. We show this in the bottom subplot of Fig. \ref{fig:fig9_barchart}. We find that shifting the zero-plane of the models improves the agreement with the data for all models other than the HMR model. The magnitude of this improvement is marginal at only 2-3\%, but we also see that the agreements above and below the zero-plane even out for the TT model and get worse for the HMR model, just as would be expected for a magnetic field structure that is antisymmetric about the plane at roughly $Z = 0.15$~kpc.

The results of this model comparison therefore suggest that the anti-symmetry of the GMF may be offset with respect to the Galactic plane. It is perhaps reasonable that the GMF would not follow the stellar plane precisely, since there are already questions raised about the extent to which it follows the spiral arms. However, this conclusion is currently limited by the number of pulsar observations available, because by shifting the parity change upwards we are transferring a lot of pulsars from being ``above'' the plane to being ``below'' the plane, which means we are losing information about the GMF above the plane.

We do not really expect any of these comparisons with analytic models to accurately represent the true structure of the GMF. This is because previous attempts \citep[e.g.][]{Noutsos2008b} have been unsuccessful in constraining model parameter values, and because our data set will necessarily include the impact of small-scale field fluctuations which are not included in the large-scale field models. This is why we use the models only as comparative indicators to test the key features of the GMF. The comparative tests done here successfully reveal some key results: that some form of sign reversals are required, and that the field parity is anti-symmetric, and that the location of the GMF plane may be offset from the stellar plane. Although we model this as being a simple vertical shift of 0.15~kpc, it is perhaps more likely that the GMF plane is tilted, or warped, which could reflect the dynamical relationship between the formation and evolution processes of both the large scale stellar and magnetic field structures of the galaxy. Investigating such structures in more depth would first require the discovery of more pulsars at larger latitudes, to fill out these regions of the map. 

The DT model does not capture the complexity of changing field directions at different distances that is observed. However, an alternative method of explaining the field reversals seen in observations would be to introduce an annular region for which the field direction is reversed. We investigate this approach in the next section.

\subsection{Testing an annular configuration}
\label{sec:annular}

Working with synchrotron observations, \cite{Vallee2022} proposes that the GMF does not maintain a consistent direction within each spiral arm, but instead has different segments pointing either clockwise or anticlockwise. The paper concludes that the GMF is predominantly clockwise, but that there exists an annular region with an anti-clockwise magnetic field lying between 5.5 and 7.6 kpc from the Galactic centre. Combining synchrotron and EGRS observations, \cite{Sun2008} determined that an axisymmetric spiral galactic field, with sign reversals introduced in a ring configuration (ASS+RING), provided the best fit to their all-sky map. In particular, they noted that a bisymmetric spiral successfully fit the data in the Galactic disk, but was not compatible alongside the halo field. Similarly, \cite{VanEck2011} described a model of a “spiralling-out” region of field reversal in an otherwise clockwise magnetic field which, given the lack of pulsars discovered at larger distances, can be viewed as qualitatively similar to a ring description for the observations available. The focus of magnetic field modelling in this paper so far has taken as a reasonable assumption that the GMF follows the stellar structure of the spiral arms, and has as a result tended towards favouring bisymmetric (or near-bisymmetric) field configurations. It is however important to test the alternative: that reversals of the GMF in an annular configuration lead to the spatially varying B field signs we observe.

We focus on testing an annular configuration qualitatively rather than quantitatively: our main goal being to see whether pulsar observations yet have sufficient density to be able to distinguish confidently between the two types of model. We therefore set up a very simple annular model to represent the key qualitative components of the various annular models described in the literature. To test this, we take the simple dipolar-toroidal model described above and modify it so that the field direction is predominantly clockwise, and symmetric above and below the plane. We then introduce an annular region into the model, from 5.5 to 7.6 kpc following \cite{Vallee2022}, in which the field direction is anticlockwise. Mathematically, we convert the field presented by \cite{Noutsos2008b} into Galactic Cartesian coordinates $X$, $Y$ and $Z$ as follows, but remove the dependence on the sign of $Z$ to make it symmetrical:
\begin{equation}
\mathbf{B} = \begin{pmatrix}\frac{2m X^2}{R^5} \pm -\frac{Y}{R}\left(\frac{m Y}{R^4} + \frac{n}{R}\right)\\\frac{2m X Y}{R^5} \pm \frac{X}{R}\left(\frac{m Y}{R^4} + \frac{n}{R}\right)\end{pmatrix},
\end{equation}
where $R = \sqrt{X^2 + Y^2}$, $m = 245~\upmu$G~kpc$^{3}$, $n = 4.8~\upmu$G~kpc and $\pm$ indicates the change of direction of the field. Then we proceed exactly as described for the previous analytic models.

\begin{figure}
    \includegraphics[width=\columnwidth]{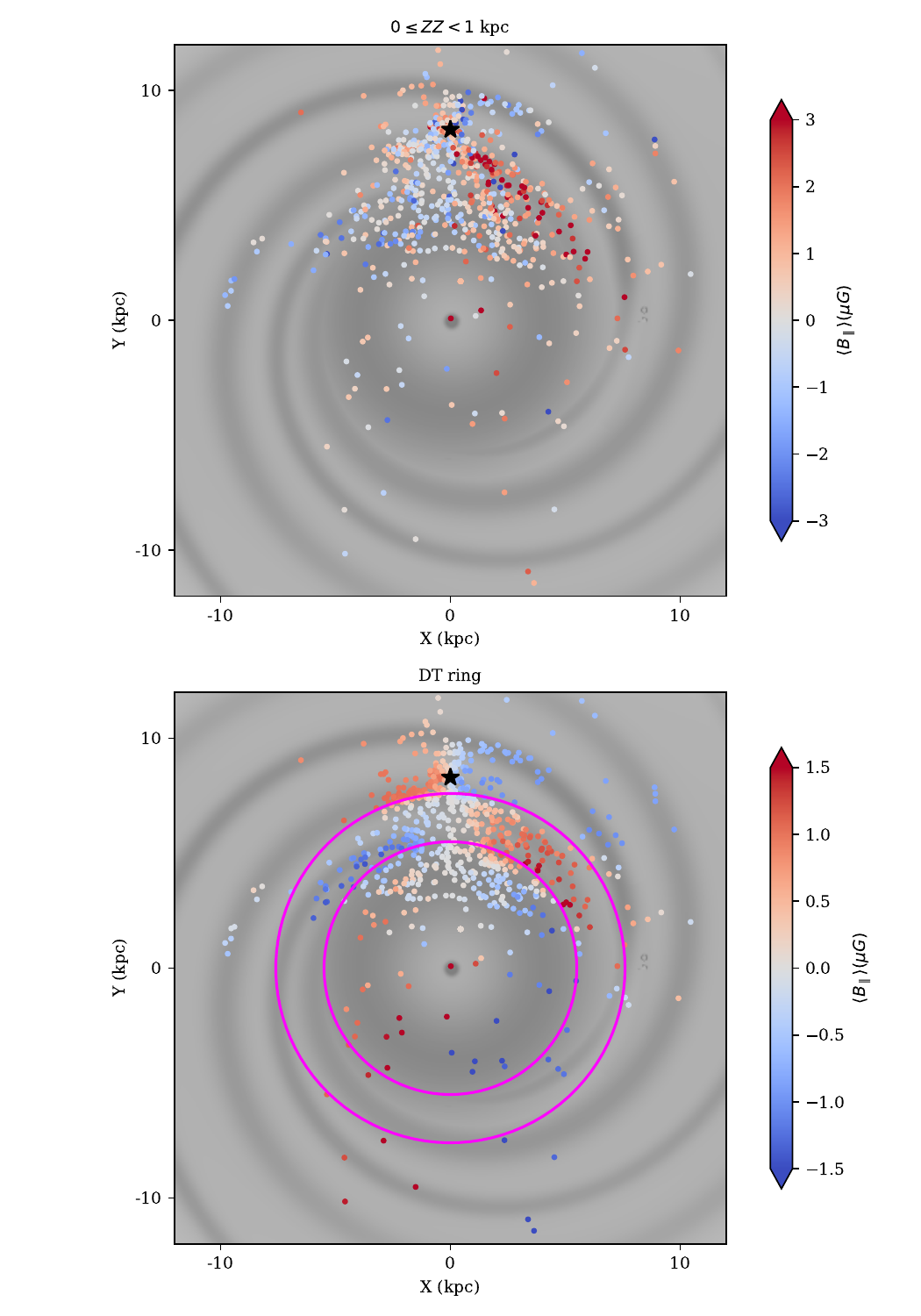}
    \caption{Top: pulsar magnetic fields in the top half of the Galactic plane ($0 \leq \lvert Z\rvert < 1$~kpc), identical to the top subfigure of Fig. \ref{fig:fig7_modelsabove}. Bottom: average magnetic field output for the DT model, adjusted to be predominantly clockwise, symmetric with respect to the Galactic plane, and with an annular region introduced in which the field direction is anticlockwise. The edges of that annular region are marked with magenta circles. Note the difference in colorbar scaling for the upper and lower figures.}
    \label{fig:fig10}
\end{figure}

\begin{figure}
    \includegraphics[width=\columnwidth]{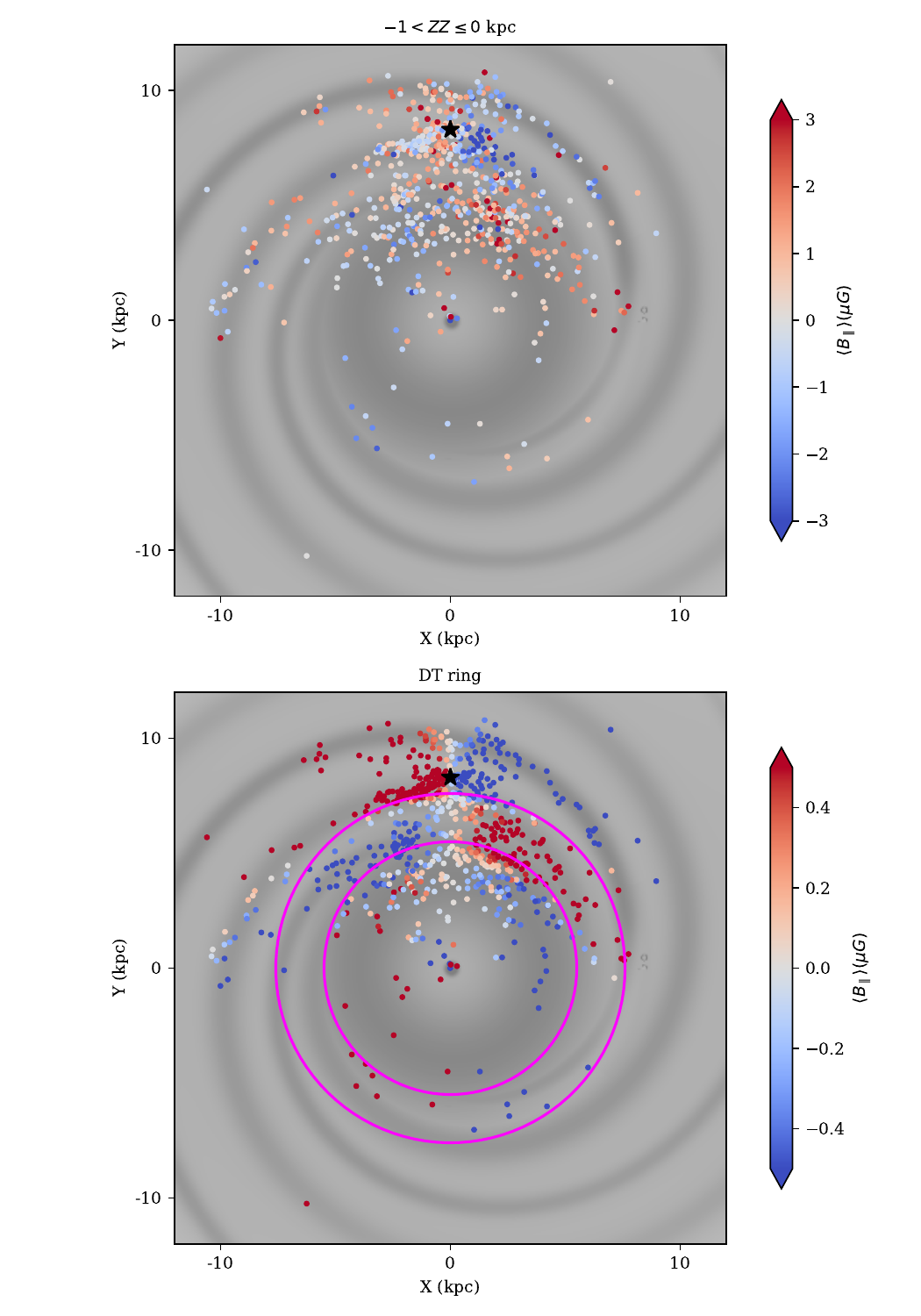}
    \caption{Top: pulsar magnetic fields in the bottom half of the Galactic plane ($-1 < \lvert Z\rvert \leq 0$~kpc), identical to the top subfigure of Fig. \ref{fig:fig8_modelsbelow}. Bottom: predicted GMF distribution in the bottom half of the Galactic disk for the DT model with an annular region, depicted as in Fig. \ref{fig:fig10}.}
    \label{fig:fig11}
\end{figure}

Plotting the resulting comparison of $\langle B_{\parallel}\rangle$ along the line of sight in Figures \ref{fig:fig10} and \ref{fig:fig11}, we see that this averaging leads, not to a simple circular shape of alternating magnetic field direction, but to a U-shaped arc, because the plotted magnetic field, $\langle B_{\parallel}\rangle$, is the average along the line of sight rather than the value at the position of the pulsar.

It should be noted that the magnetic field strengths, for this configuration of the ring model, are much weaker than those observed (note that the colorbars in Figures \ref{fig:fig10} and \ref{fig:fig11} are scaled differently to all other colorbars in this paper). This has happened because we have introduce a sign change to the dipolar-toroidal model used earlier, without making any adjustments to the model magnetic field strength scaling. This will lead to cancelling out of magnetic fields along the lines of sight where the ring sign changes are included. This is not a factor of concern in comparing this model with the data, because our focus is on the sign of the magnetic field, not its magnitude. Visually, the distribution of positive and negative $\langle B_{\parallel}\rangle$ measurements is roughly as plausible as those for the antisymmetric logarithmic spiral models: there are variations in the precise locations in the sign changes of the $\langle B_{\parallel}\rangle$ but observational fidelity is insufficient to discriminate between the different models as yet. Previous comparisons of spiral models and concentric ring models by \cite{Han1999} and \cite{Indrani1999} both concluded that a bisymmetric spiral fit the observed data better than a concentric ring structure. \cite{Vallee2022} still favoured a spiral model, but was focused on the sign changes of the field directions following an annular region, rather than alternating between the arm and inter-arm regions as preferred by \cite{Han2018}, who worked with pulsar observations. \cite{Sun2008} again favoured a spiral model to explain synchrotron and EGRS observations, with the annular region conferring only the necessary change in field direction to introduce observed sign reversals, and \cite{VanEck2011} found the same, again for synchrotron and EGRS observations. Our results demonstrate that we are still not yet in a regime where we are able to determine confidently the accuracy of the locations of direction changes in the GMF. Although pulsars provide an increased coverage of the three-dimensional disk field distribution than the average effect of EGRS, like \cite{Unger2023} we find that the idea that the GMF in the plane follows a large-scale spiral structure cannot be concluded with absolute certainty. The increased measurement precision highlights the constraints of our limited knowledge of pulsar distances, and additional complexities of the magnetic field that are not well captured statistically by either type of model.


\section{Discussion}
\label{sec:discussion}

Using pulsars to study the magnetic field of the Milky Way has a unique set of advantages and disadvantages. Pulsars are unique in both providing a direct measurement of the average magnetic field component along the line of sight and giving a three-dimensional perspective of the GMF. However, this measurement is of the cumulative effect along the whole line of sight, which blurs our understanding, particularly as we look out to greater distances, where the two issues of a lack of pulsars discovered, and the increasing likelihood of averaging over multiple field reversals, combine to blur the picture. Furthermore, the fact that we do not have independent distance measurements to the majority of pulsars is a key limitation on our ability to interpret $\langle B_{\parallel}\rangle$. In particular, it is likely to affect how well we can distinguish between positions on the spiral arms and positions between the spiral arms, which is strongly relevant for determining the relationship of the magnetic field structure to the stellar structure of the galaxy. 

Despite these limitations, a considerable amount may be learned from pulsar measurements of the GMF, particularly now that we have access to 1,838 lines of sight to map out the sky. This work supports previous interpretations that have suggested that the GMF is best described by a (roughly) bisymmetric logarithmic spiral structure which is antisymmetric with respect to the plane, and comparison with extragalactic radio sources supports our interpretation. \cite{Nota2010} pointed out that, for their study of the fourth Galactic quadrant, ``the details of the spiral arm model... hardly affect the character of the solution''. The same appears true when considering the whole galaxy, as we do here: comparing the TT, HMR and PS models it can be seen that the choice of pitch angle has little influence on how well the model reproduces the data. It is, alternatively, still potentially possible to use an annular region to model much of the GMF sign reversals, perhaps retaining a spiral structure which is axisymmetric and introducing an annular region to model the reversals which are otherwise attributed to bisymmetry, but there is insufficient evidence to promote this particular model as preferable to a bisymmetric spiral structure. 

We note the work of \cite{West2021}, which used the Planck map of linear polarized intensity \citep{Adam2016} to study the Northern Polar Spur and Fan Region. They concluded that both observed structures could be modelled as resulting from long, parallel, magnetized filaments pointing along the Local arm (also known as the Orion Arm, Orion-Cygnus Arm or Orion Spur, which is how it is labelled in Fig. \ref{fig:Bfieldinplane}). Similarly, \cite{Hutschenreuter2020} found evidence that the magnetic field is aligned with the Orion arm. The presence of such local magnetized filaments will impact the magnetic field observed along the line of sight of these filaments. \cite{West2021} find that an orientation towards Galactic longitude $l = 45\degree$ is a reasonable match to their observations, while \cite{Hutschenreuter2020} indicate a pronounced structure in their maps at $l = 60\degree$. We note that the region with the strongest magnetic field measurements in our own data set lies in the Sagittarius arm between $40\degree$ and $50\degree$ of longitude, but that, closer to the Sun and above the plane, the magnetic field direction points away from the Sun rather than towards it. The sudden change in sign with increasing difference could therefore be attributed to a change from the filaments of the Local arm dominating the observations, to the Sagittarius arm field having the strongest effect on observations of more distant pulsars. The filamentary structures inferred by \cite{West2021} may have similar counterparts throughout the rest of the galaxy: these may be the cause of the observation that the magnetic field seems to trace the spiral arms. 

We also note the work of \cite{Ordog2017}, who compared Galactic diffuse emission and EGRS RMs to find a diagonal RM gradient associated with the field reversal in the Sagittarius-Carina arm. They interpreted this as the GMF structure containing a current sheet that is not perpendicular to the Galactic plane, leading to the diagonal reversal. They comment on the potential impacts of undulating field reversals originating from convective instabilities, and a potential misalignment between the Galactic plane and its magnetic dipole. These interpretations align with our comments about both the incomplete antisymmetry of the GMF about the plane, depending on the spiral arms, and our potential misalignment of the Galactic and stellar planes, as discussed in sections \ref{sec:spiraltests} and \ref{subsec:comparemodels}. The specific location of their field reversal ($52\degree<l<72\degree$) is also worthy of closer study in the light of our own comparisons to EGRS at $60\degree<l<70\degree$, which suggested a special local field configuration at an 8 kpc distance from the Sun and above the plane only. At closer distances to the Sun in this region we also see strong antisymmetry of B field signs with respect to the plane, which agrees qualitatively with the field reversal gradient discussed by \cite{Ordog2017}. We can also compare the work of \cite{Ma2020}, who proposed, from comparison of their new EGRS measurements with pulsar RMs, that the Sagittarius arm hosts an odd-parity disc field. Our own findings provide good support for such a model describing this region. Finally, we note the work by \cite{West2021} discussed above, relating to the impact of the Local Arm on GMF measurements around $l=45\degree$. A more detailed three-dimensional study, relating RM gradients in relation to the plane, with the regions of interest at different distances commented on here, would be able to constrain our understanding of this region of the sky, and enable us to map out the full physical impact of field reversals throughout the Galaxy.

The increased fidelity of observations has enabled us to investigate the plane antisymmetry more closely, showing that the best fitting models place the GMF plane of antisymmetry at $Z \approx 0.15$~kpc. In addition, the magnetic field directions along each of the spiral arms is not antisymmetric with respect to the plane in all cases (notably the Perseus arm). Both of these features raise interesting questions about the relationship of the GMF to the spiral arms in terms of their formation history. The existence of field reversals in the Milky Way is unusual, perhaps unique, in the context of magnetic field observations of other galaxies \citep{Beck2016}, and Amp\`ere's law implies the presence of a current sheet at the location of each reversal, the origins of which are unclear. However, we are increasingly seeing more complex magnetic field configurations for other galaxies as well, such as the extreme pitch angle of the magnetic field vectors of NGC 2997 (A. Damas Segovia, private communication and paper in prep.). Further high fidelity observations of Galactic field structures will surely continue to open up new challenges to theories of magnetic field formation.

Many previous publications have discussed the presence of sign reversals of the GMF between arm and interarm regions, using both pulsar measurements of the parallel field component \citep{Han2018} and synchrotron tracers of the perpendicular field component \citep{Vallee2022}. We note that by separately considering the magnetic field above and below the plane, and by adjusting the position at which we consider the split, we reduce some of the need to introduce reversals along the line of sight by instead accounting for it as field antisymmetry above and below the plane. We are cautious about introducing arm/interarm sign reversals into either our simple spiral picture or the analytic magnetic field models, due to the limitations on our knowledge of distances to pulsars, and instead investigate the extent to which we can explain some of the sign reversals in regions close to the Sun as resulting from the impact of localised effects in certain regions, particularly the effect of filamentary structures in the Local arm, rather than the large-scale GMF. In general, the GMF is likely to be tied to the structures of the spiral arms themselves, which may include inclinations and warps. Future GMF modelling may benefit from including more detailed spiral arm structure information, such as the age patterns and the Galaxy corotation radius \citep[e.g.][]{He2021}, and, conversely, GMF results may be able to inform such Galactic structure models.

\section{Conclusions}
\label{sec:conclusions}

Using 19,697 observations of 1,097 pulsars from the Thousand-Pulsar-Array survey, combined with existing measurements to 741 pulsars from the pulsar catalogue, we have formed the most comprehensive map to date of pulsar observations of the Galactic magnetic field. Comparing these measurements to the rotation measures from extragalactic radio sources, and to models of the GMF that include one simple spiral picture, one analytic dipolar-toroidal model, three analytic logarithmic spiral models, and one annular ring-shaped model, we find the following results. 

The average magnetic field estimates inferred from pulsar measurements correspond well to the idea that the GMF follows the spiral arms of the galaxy. This approximately resembles a roughly bisymmetric spiral structure. The varying strength of the magnetic field matches well with the directions of the spiral arms relative to our vantage point on Earth, and comparisons of sign changes in the magnetic field with EGRS back up this interpretation. Complications to this picture can largely be explained by the GMF being predominantly antisymmetric with respect to the Galactic plane, but with some spiral arms being symmetric instead. More detailed modelling suggests that this plane anti-symmetry is offset from the stellar plane. Indeed, the plane of anti-symmetry could even be tilted or curved, which might account for the varying symmetries/anti-symmetries of different spiral arms, but there are currently insufficient measurements to pulsars to investigate this hypothesis more closely. 

In this study we have focused on using pulsars and EGRS as probes of the GMF in the Galactic disk. Future work would benefit from extending this study into the wider context, both of observations and of the disk in the context of the halo field. The number and precision of the measurements making up this data set represent a sizable advance in our capacity to study the GMF using pulsars, but a key limitation remains in our lack of precise knowledge of the distances to pulsars, particularly those further away from Earth, and our consequent heavy reliance on modelling the electron column density to be able to draw conclusions about the three-dimensional structure of the GMF. Future advances will be possible with more in depth comparison of pulsar measurements of the GMF with those obtained through alternative observations, such as those of EGRS \citep{Hutschenreuter2023}, synchrotron emission \citep{West2021}, dust polarization \citep{Pelgrims2021}, starlight polarization \citep{Pelgrims2024} and diffuse galactic emission \citep[e.g.][]{Ordog2019}, and of discovery of new pulsars with the Square Kilometre Array telescopes. This will increase the number of lines of sight through the ISM, particularly to fainter, more distant pulsars, advancing our ability to probe the GMF beyond our local region of the galaxy.

\section*{Acknowledgements}
The MeerKAT telescope is operated by the South African Radio Astronomy Observatory (SARAO), which is a facility of the National Research Foundation, an agency of the Department of Science and Innovation. SARAO acknowledges the ongoing advice and calibration of GPS systems by the National Metrology Institute of South Africa (NMISA) and the time space reference systems department department of the Paris Observatory. 
PTUSE was developed with support from the Australian SKA Office and Swinburne University of Technology. 
This work made use of the OzSTAR national HPC facility at Swinburne University of Technology.
MeerTime data is housed on the OzSTAR supercomputer.
The OzSTAR program receives funding in part from the Astronomy National Collaborative Research Infrastructure Strategy (NCRIS) allocation provided by the Australian Government.
LSO acknowledges the support of Magdalen College, Oxford. 
Pulsar research at Jodrell Bank Centre for Astrophysics and Jodrell Bank Observatory is supported by a consolidated grant (ST/X001229/1) from the UK Science and Technology Facilities Council (STFC).
We acknowledge the use of the following Python packages: 
\texttt{MATPLOTLIB} \citep{Hunter2007} and
\texttt{PANDAS} \citep{McKinney2010}.

\section*{Data Availability}
The data underlying this article are published by \cite{Posselt2023} and \cite{Keith2024} and the RM and other associated measurements are all presented in the Supplementary Material for ease of further analysis and as a contribution to catalogues of galactic RMs.




\bibliographystyle{mnras}
\bibliography{Magnetism} 


\section*{Supplementary material}
Full, machine-readable versions of Tables \ref{tab:RMvaluestablesubset} and \ref{tab:RMDMdistBvaluestablesubset} are available at MNRAS online. The results are also formatted in the style of the RMTable catalogue for easy integration, and these may also be found at MNRAS online.

\appendix
\section{Mathematics of analytic models}
\label{appendix:ModelMaths}

\noindent\textbf{DT model:}
\begin{equation}
\mathbf{B} = \begin{pmatrix}\frac{m}{r^5}(2X^2 - Y^2) - \textrm{sign}(Z)\frac{nY}{r^2}\\\frac{3m X Y}{r^5} + \textrm{sign}(Z)\frac{nX}{r^2}\end{pmatrix},
\end{equation}
where $r = \sqrt{X^2 + Y^2}$, $m = 245~\upmu$G~kpc$^{3}$ and $n = 4.8~\upmu$G~kpc.\\

\noindent\textbf{TT model:}
\begin{equation}
\mathbf{B} = \begin{pmatrix}(A/r)X\sin(p) - Y\cos(p)\\(A/r)Y\sin(p) + X\cos(p)\end{pmatrix},
\end{equation}
where $r$ and $\theta$ are standard polar coordinates derived from $X$ and $Y$,\\ 
$A = b\cos(\theta - \ln(r/r_0)/\tan(p))f_z$,\\ 
$r_0 = (r_{\bigodot} + d_0)\exp^{-0.5\pi\tan(p)}$,\\ 
$f_z = \textrm{sign}(z_0)\exp(-\lvert Z\rvert/Z_0)$,\\ 
\[
b=\begin{cases}
B(r_{\bigodot})r_{\bigodot}/(r\cos(\phi)), & \text{if $r > r_{core}$}\\
B(r_{\bigodot})r_{\bigodot}/(r_{core}\cos(\phi)), & \text{otherwise}
\end{cases}
\]\\
$\phi = (1/\tan(p))\ln(1 + d_0/r_{\bigodot}) - \pi/2$\\ 
$p = -8$ radians, $d_0 = -0.5$ kpc, $r_{core} = 4$ kpc, $Z_0 = 1.5$ kpc, and $B(r_{\bigodot}) = 1.4 \upmu$G.\\

\noindent\textbf{HMR model:}

\noindent As for the TT model, but now \\
$f_z = 1/(2\cosh(Z/z_1)) + 1/(2\cosh(Z/z_2))$,\\ 
where $z_1 = 0.3$ kpc and $z_2 = 4$ kpc,\\ 
$b = 3(r_{\bigodot}/r)(\tanh(r/r_{core}))^3$,\\
$p = -10$ radians, $d_0 = -0.5023$ kpc, and $r_{core} = 2$ kpc.\\

\noindent\textbf{PS model:}\\
\noindent This model consists of three components. The first is the same as for the TT model, except\\ 
$f_z = \exp(-\lvert Z\rvert)/Z_0)$, \\
$p = -8$ radians, 
$d0 = -0.5$ kpc, 
$r_core = 4$ kpc, 
$Z0 = 0.2$ kpc, 
$B_rsun = 2 \upmu$G.\\

\noindent Next is a toroidal contribution:\\
$B_{T, max}(r_{\bigodot}) = 1.5 \upmu$G, 
$hT = 1.5$kpc, 
$wT = 0.06$kpc, \\
$Hfunc = (H(r_{\bigodot}-r, 1) + H(r-r_{\bigodot}, 1)\exp((r_{\bigodot}-r)/r_{\bigodot}))/(1 + (\lvert Z\rvert-h_T)/w_T)^2)$\\
\begin{equation}
\mathbf{B_T} = \begin{pmatrix} -\textrm{sign}(Z)B_{T, max}(r_{\bigodot})(Hfunc)\cos(\theta) \\ \textrm{sign}(Z)B_{T, max}(r_{\bigodot})(Hfunc)\sin(\theta) \end{pmatrix},
\end{equation}

\noindent Finally a dipole component, given in terms of spherical polar coordinates $r$, $\theta$ and $\phi$:\\
\begin{equation}
\mathbf{B_D} = \begin{pmatrix} -3\frac{\upmu_{G}}{R^3}\cos{\phi}\sin{\phi}\sin{\theta} \\ -3\frac{\upmu_{G}}{R^3}\cos{\phi}\sin{\phi}\cos{\theta} \\ \frac{\upmu_{G}}{R^3}(1 - 3\cos^{2}{\theta})) \end{pmatrix},
\end{equation}
where $\upmu_{G} = 123 \upmu$G~kpc$^{3}$, and the z-component of the field, $B_{z} = -100 \upmu$G for $r < 500$ pc.\\

\bsp	
\label{lastpage}
\end{document}